\documentclass[conference]{IEEEtran}

\usepackage{booktabs} 

\usepackage{fancyhdr}
\usepackage[normalem]{ulem}
\usepackage{hyperref}
\hypersetup{hidelinks}
\usepackage{CJK}
\usepackage{graphicx}
\usepackage{epstopdf}
\usepackage{subfigure}
\usepackage{xcolor}
\usepackage{colortbl}
\usepackage{lipsum}
\usepackage{multirow}
\usepackage{booktabs}
\usepackage{tabularx}
\usepackage{bbding}
\usepackage{color}
\usepackage{caption}
\usepackage{amsmath}
\usepackage{layout}
\usepackage{algorithm}
\usepackage[noend]{algpseudocode}
\usepackage{algcompatible}

\usepackage{enumitem}
\usepackage{eqparbox}

\newcommand{\ignore}[1]{}
\newcommand{\tabincell}[2]{\begin{tabular}{@{}#1@{}}#2\end{tabular}}

\makeatletter
\def\@copyrightspace{\relax}
\makeatother
\ifCLASSOPTIONcompsoc
  \usepackage[nocompress]{cite}
\else
  \usepackage{cite}
\fi
\author{\IEEEauthorblockN{Xiaoyuan Wang}
\IEEEauthorblockA{Huazhong University of Science and Technology\\
xiaoyuanw@hust.edu.cn}
}

\title{Supporting Superpages and Lightweight Page Migration in Hybrid Memory Systems}

\begin{document}
\maketitle


\begin{abstract}

Superpages have long been used to mitigate address translation overhead in big memory systems. However, superpages often preclude lightweight page migration, which is crucial for performance and energy efficiency in hybrid memory systems composed of DRAM and non-volatile memory (NVM). In this paper, we propose a novel memory management mechanism called \textit{Rainbow} to bridge this fundamental conflict between superpages and lightweight page migration. \textit{Rainbow} manages NVM at the superpage granularity, and uses DRAM to cache frequently-accessed (hot) small pages in each superpage. Correspondingly, \textit{Rainbow} utilizes split TLBs to support different page sizes. By introducing an efficient hot page identification mechanism and a novel NVM-to-DRAM address remapping mechanism, \textit{Rainbow} supports lightweight page migration while without splintering superpages. Experimental results show that Rainbow can significantly reduce applications' TLB misses by 99.8\%, and improve application performance (IPC) by up to 2.9X (43.0\% on average) when compared to a state-of-the-art memory migration policy without superpage support.

\end{abstract}
\IEEEpeerreviewmaketitle
\section{Introduction}\label{Sec:intro}

Today's data-intensive applications like big data processing, live streaming~\cite{anysee} and graph analytics place heavy demands on memory capacity. DRAM scaling, however, is unable to fit the increasing memory requirement for petabyte-scale big data processing. Emerging byte-addressable non-volatile memory (NVM) technologies, such as phase change memory (PCM) and 3D XPoint~\cite{3DXPoint} offer high memory density, low cost per bit and near-zero standby power consumption, at the expense of low performance and limited write endurance~\cite{dhiman2009pdram,Qureshi:2009}. Despite the promising features of NVM, it is not exactly a direct substitute for DRAM. Thus, it is more practical to use NVM in conjunction with DRAM in hybrid memory systems~\cite{Qureshi:2009,dhiman2009pdram,ramos2011page,wei2015exploiting,Liu:2017}.




With the continuously increasing application footprints and a corresponding growth of memory capacity, virtual-to-physical memory address translation tends to be a new bottleneck of system performance~\cite{Alam:2017:DVM:3079856.3080209}. Modern computer systems typically employ translation lookaside buffer (TLB) as a cache to store the recent virtual-to-physical address translations for faster retrieval in the future. Upon each memory reference, the TLB is consulted first. If the requested address is not present in the TLB (i.e., a TLB miss), the CPU needs to retrieve the absent address translation through hardware page table walking, which incurs a significant performance penalty due to four memory references in x86-64 systems~\cite{Yaniv:2016:HDC:2896377.2901456}. Previous studies have shown that TLB misses may significantly degrade system performance by up to 50\% when the application's memory footprint becomes very large~\cite{mccurdy2008investigating,basu2013efficient,Barr:2010:TCS:1815961.1815970,Bhattacharjee:2013:LMM:2540708.2540741,du2015supporting}. 

There has been a large body of work on improving TLB coverage, i.e., the total memory address space that can be directly translated through TLBs. As the number of TLB entries does not scale up well due to speed, power, and space constraints, superpages have been widely exploited to increase TLB coverage~\cite{du2015supporting,Swanson:1998,Romer:1995,Cox:2017,Agarwal:2017}. A superpage refers to a large virtual page that maps to a number of continuous physical small (base) pages. The use of superpages can significantly broaden the TLB coverage (by a factor of 512 for typical 2 MB superpages compared to 4 KB small pages). However, the side effect is that superpage can hamper lightweight and agile memory management, such as page migration. 



On the other hand, hybrid DRAM/NVM memory systems often rely on page migrations to achieve higher performance~\cite{Qureshi:2009,ramos2011page,wei2015exploiting,Liu:2017} and energy efficiency~\cite{park2011power,wei2015exploiting,Salkhordeh:2016,Liu:2017}. This in turn requires lightweight page migration schemes to move the frequently-accessed (hot) pages from the slow NVM to the fast DRAM. However, page migration at the superpage granularity (e.g., 2 MB) can incur unbearable performance overhead due to a vast waste of DRAM capacity and bandwidth if most memory references are distributed in a small region of the superpage (see Section~\ref{Sec:observation}). The cost may be even larger than the benefit of superpage migration. This places the use of superpage in a dilemma since the lightweight page migration can outweigh the benefits of extended TLB coverage.

In this paper, we study how to exploit superpages for wide TLB coverage while supporting lightweight page migration in hybrid memory systems. To achieve this goal, several challenging issues should be addressed. (1) \textbf{Lightweight hot pages identification:} to support lightweight page migration, a large body of work advocates monitoring memory accesses through the memory controller~\cite{park2011power,ramos2011page}. However, access counters at per-page granularity (i.e., 4 KB) leads to prohibitively high storage overhead when the capacity of main memory becomes large (e.g., 1 TB memory needs total $\frac{1TB}{4KB}\times 2B = 512 MB$ storage on a 2B-per-page basis). Storing those records with on-chip SRAM is impractical. Another alternative is to store them in main memory, however, this would lead to higher memory access latency and additional records lookup overhead in main memory for each memory reference. (2) \textbf{Impact of lightweight page migration on TLB coverage}: page migrations often fragment superpages and thus break the physical address continuity. Previous work has advocated splintering superpages to enable lightweight memory management such as page migration and sharing, while sacrificing the performance of address translation~\cite{Pham:2015,Guo:2015}. It is still a challenge to retain the improved TLB coverage when the hot small pages within superpages are migrated to the DRAM. (3) \textbf{Efficiency of hot pages addressing}: as hot pages contribute to a major portion of applications' memory references, it is essential to further improve address translation performance for those hot pages in the DRAM. 

To address the above problems, we propose \textit{Rainbow}, a novel memory management mechanism to bridge the gap between superpaging and lightweight page migration for hybrid DRAM/NVM memory systems. \textit{Rainbow} manages NVM and DRAM with different page granularities. Correspondingly, \textit{Rainbow} exploits the available hardware feature of split TLBs ~\cite{papadopoulou2015prediction,Cox:2017,AMDFamily,IntelSkylake} to support different page sizes, with one TLB for addressing the superpages, and another TLB for small pages. Rainbow migrates hot small pages within superpages to the DRAM, without compromising the integrity of superpage TLB. As a result, Rainbow actually architects the DRAM as a cache to the NVM.
Rainbow has the following novel designs to address the aforementioned challenging issues in supporting both superpages and lightweight page migration:

\begin{itemize}[leftmargin=*]
\item
To mitigate the storage overhead of fine-grained page access counting, we propose two-stage memory access counting. In a given time interval, Rainbow first counts NVM memory accesses at the superpage granularity, and then selects the $top$ $N$ hot superpages as targets. At the second stage, we only monitor those hot superpages at the small pages (4 KB) granularity to identify hot small pages. This history-based policy avoids monitoring the sub-blocks (4 KB pages) within a large number of cold superpages, and thus significantly reduce the overhead of hot page identification.
\item
We adopt split TLBs to accelerate the address translation performance for both DRAM and NVM references. To keep the integrity of superpages' TLB when some small pages are migrated to the DRAM, we use a bitmap to identify the migrated hot pages in the memory controller, without splintering the superpages.
\item
We propose a physical address remapping mechanism to access the migrated hot pages in the DRAM, without suffering costly page table walks for addressing a DRAM page. To achieve this goal, we store the migrated hot page's destination address in its original residence (the superpage). Once the hot pages' corresponding TLB misses, the DRAM page addressing should resort to an indirect access of the superpage. This design logically leverages the superpage TLBs as a next-level cache of the 4 KB page TLBs. Because the superpage TLB hit rate is rather high, Rainbow can significantly speed up the DRAM page addressing.


\end{itemize}

Putting those designs all together, we implement Rainbow within an integrated simulator based on zsim~\cite{Sanchez:2013} and NVMain~\cite{NVMain}. To the best of our knowledge, this is the first kind of work that supports superpages and lightweight page migration in hybrid memory systems. We compare Rainbow with several alternatives using a wide range of workloads. Experimental results show that Rainbow can significantly reduce the address translation overhead for applications with large memory footprints, and thus improve application performance by up to 2.9X (43.0\% on average) compared to a hybrid memory migration policy without superpage support. Rainbow also demonstrates higher energy efficiency than other policies.

The remainder of this paper is organized as follows. Section~\ref{sec:motivation} introduces the background and motivates our design for DRAM/NVM hybrid memories. Section~\ref{sec:design-and-impl} gives the detailed design of Rainbow.  Experimental results are presented in Section~\ref{sec:evaluation}. We discuss the related work in Section~\ref{relatedwork} and conclude in Section~\ref{sec:conc}.


%
%
%
%

\section{Background and Motivation} \label{sec:motivation}
We first introduce the superpages and split TLBs. Next, we experimentally study memory access statistics of typical applications to motivate the design of Rainbow.
%
%
%
\subsection{Superpage and Split TLBs}
The evidence of performance degradation due to address translation have been well corroborated~\cite{mccurdy2008investigating,basu2013efficient,Barr:2010:TCS:1815961.1815970,Bhattacharjee:2013:LMM:2540708.2540741}. Modern data-centric applications are characterized with large memory footprint and lower data locality, and are expected to incur even higher address translation overheads due to TLB misses. On the other hand, emerging NVM technologies are much denser and cheaper than the conventional DRAM, and consequently we expect a rapid growth of memory capacity in the near future. The trends of big memory systems make the address translation problem become more urgent than ever before.

TLB misses can be mitigated by improving TLB coverage (or TLB reach). There are two fundamental ways to enlarge the TLB coverage, either by using more TLB entries or letting each entry map a larger memory page. Increasing TLB size implies larger die space area, higher energy consumption and access latency. Another alternative is to use superpages, which have been proposed to improve TLB coverage since the 1990s~\cite{Talluri:1994,Romer:1995}. Most modern computer systems support superpages at both hardware and software levels. For example, x86-64 supports 4 KB, 2 MB and 1 GB page sizes, and the processor vendors provide split TLBs for different page sizes correspondingly~\cite{papadopoulou2015prediction,Cox:2017,AMDFamily,IntelSkylake}.  A virtual address can be consulted in all split TLBs in parallel to shorten the address translation latency.  Although split TLBs are simple for implementation and with good performance, they would be underutilized without judicious allocation of superpages at different sizes. For example, if the OS only allocates 4 KB pages, the 2 MB superpage TLBs are wasted.


%
%
%
%

\subsection{Memory Access Analytics of Superpages}\label{Sec:observation}

Emerging NVM offers higher density than DRAM, but at the expense of higher access latency and lower bandwidth. Particularly, for the write operations, NVM is about 5-10x slower than DRAM, and consumes up to 10x more energy than DRAM~\cite{Raoux:2008}. As a result, page migration is widely utilized to improve performance and energy efficiency in hybrid memory systems~\cite{Qureshi:2009,park2011power,ramos2011page,wei2015exploiting,Liu:2017}. However, the use of superpages in hybrid memory systems precludes lightweight page migrations because
a superpage is required to be contiguous and aligned in both physical and virtual address spaces. Fine-grained page (e.g., 4 KB) migration breaks the continuity of physical address space, and thus splinters the superpages. Page migration at the superpage granularity (e.g., 2 MB) can still retain the advantages of wide TLB coverage, however, is prohibitively costly.

To evaluate the side effects of superpage migrations, we run several representative applications using 2 MB superpages, and profile their memory usage at the granularity of 4 KB pages in an interval of $10^8$ cycles. These applications are selected from SPEC CPU2006\cite{Spec_CPU2006}, Parsec\cite{Parsec}, Problem Based Benchmarks Suit (PBBS)\cite{PBBS}, Graph500\cite{Graph500},  Linpack\cite{Linpack}, NPB-CG~\cite{CG}, and HPC Challenge Benchmark GUPS~\cite{GUPS}. CactusADM, mcf and soplex are selected from SPEC CPU2006. CactusADM is a computational kernel representative of many applications in numerical relativity. Soplex solves a linear program using the simplex algorithm. Mcf is a program used for single-depot vehicle scheduling in public mass transportation. Canneal, bodytrack and streamcluster are multi-thread applications selected from Parsec. DICT, BFS, setCover and MST are selected from PBBS. Both BFS and MSF all solve graph problems. SetCover is a computational biological problem. DICT is a dictionary matching algorithm.  Linpack is a traditional supercomputer benchmark performing numerical linear algebra. Graph500 is a supercomputer benchmark based on large-scale data-intensive graph analysis.  NPB-CG measures irregular memory access and communication performance. GUPS measures the rate of integer random updates of memory. These workloads cover a wide range of memory access patterns, and their memory footprints are shown in Table~\ref{tab:hotpage}. All experiments are conducted in a simulated platform, as presented in Section~\ref{test-setup}. We have the following observations.

\begin{figure}[tbp]
 \vspace{-0.0ex}
\centering
\includegraphics[width=0.9\columnwidth]{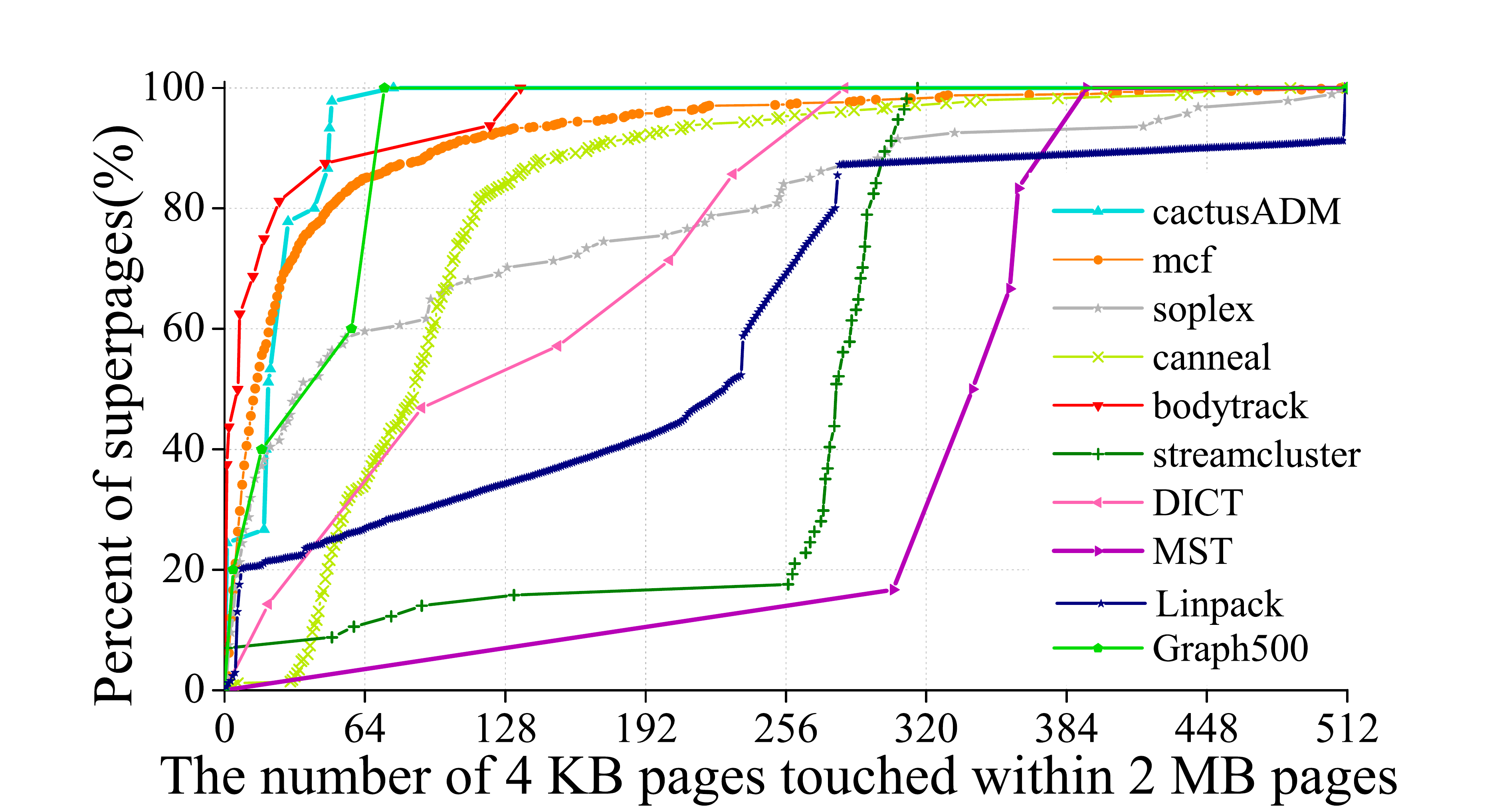}
\vspace{-0.5ex}
\caption{The cumulative distribution function of superpages versus the number of touched 4 KB small pages within a superpage in a given interval}
\label{fig:memory_footprint}
 \vspace{-1ex}
\end{figure}

\emph{Observation 1: For most applications, only a small portion of 4 KB pages are actually touched in each superpage during each sampling interval.} Figure~\ref{fig:memory_footprint} shows the cumulative distribution function (CDF) of superpages as a fraction of the touched small pages in one superpage. For many applications, we find that almost 80\% superpages' memory accesses are distributed on only a few small pages in a given interval. For cactusADM, the total number of touched small pages is even less than 100 for all superpages. This indicates that the migration of a whole superpage often results in wasted DRAM bandwidth and CPU time, and inefficient use of the limited DRAM capacity. The cost may be even larger than the benefit of superpage migration.


\emph{Observation 2: most applications' memory references are mainly distributed within a small portion of 4 KB hot pages.} Similar to CHOP~\cite{jiang2010chop},
the hot pages are classified as the top N pages ranked by number of accesses, and the total accesses of these pages constitute 70\% of the application's memory accesses. For each application, Table~\ref{tab:hotpage} shows the minimum access counts for hot pages in working sets measured every $10^8$ cycles, and the applications' total memory footprints. The hot page percent is calculated as the total volume of small hot pages to the working set in the sampling interval. Given the small fraction of touched small pages in each superpage, the proportion of hot small pages is even much smaller for many applications, such as mcf, canneal, and bodytrack. Table~\ref{tab:hotsmallpage} shows how the hot small pages are distributed among superpages. For many applications, we find that most superpages' memory references are distributed on less than 128 hot small pages. This is extremely clear for data-intensive benchmarks. For GUPS and Graph500, even 95.5\% and 61.48\% superpages are covered by less than 32 hot small pages. This implies it is more beneficial and lightweight to migrate only the hot small pages rather than the whole superpages from NVM to DRAM.

\begin{table}[tbp]
\scriptsize
\centering
\caption{ Hot Page (4 KB) Access Statistics}
\label{tab:hotpage}
\vspace{-0ex}
\begin{tabular}{|c|c|c|c|c|c|}
\hline
\multirow{3}{1.5cm}{Application} &  \multicolumn{3}{c|}{\tabincell{c}{{Page access statistics} ($10^8$ cycles)}} & \multirow{2}{1.1cm}{{Total memory footprint}}  \\
\cline{2-4}
    & \tabincell{c}{Hot page\\ min\# access}
    &\tabincell{c}{Working\\ set (MB)}&
    \tabincell{c}{Hot page\\percent} &\\
\hline
cactusADM        &64  & 74.6 MB &  4.71\% & 776 MB  \\
mcf              &30  & 1089 MB &  2.36\% & 1698 MB \\
soplex           &51  & 70.9 MB & 19.63\% & 1888 MB  \\
\hline
canneal          &2  & 891.6 MB &  8.52\%& 972 MB  \\
bodytrack        &19  & 16.2 MB &   1\%   &  620 MB  \\
streamcluster    &10  & 105.5 MB & 27.60\% &150 MB \\
\hline
DICT             &53  &  20.3 MB& 37.20\% &384 MB  \\
BFS              &30  & 404.1 MB& 20.51\% &3718 MB  \\
setCover         &34  &  49.8 MB& 37.53\% &2520 MB  \\
MST              &35  & 121.2 MB& 32.42\% &6660 MB  \\
\hline
Graph500       &64  & 7.20 MB &  6.35\% & 27.4 GB  \\
\hline
Linpack         &63   &40 MB  & 21.19\%  & 23.9 GB \\
\hline
NPB-CG         &64   &40.9 MB  & 24.7\%  & 22.9 GB \\
\hline
GUPS            &4 &7.6 GB  & 5.8\%  & 8.06 GB \\
\hline
\end{tabular}
\vspace{-1ex}
\end{table}




The above observations motivate us to design a new memory management mechanism that supports both superpages and fine-grained page migration for hybrid memory systems.

\begin{table}[tbp]
\scriptsize
\centering
\caption{ Distribution of Hot 4 KB Pages within Superpages }
\label{tab:hotsmallpage}
\begin{tabular}{|c|c|c|c|c|c|c|}
\hline
\multirow{2}{1cm}{Application} & \multicolumn{6}{c|}{Percent of superpages covered by a number of hot 4 KB pages } \\
\cline{2-7}
 &1-32 &33-64 &65-128 &129-256 &257-384 &385-512 \\
  \hline
cactusADM       &28.01\%	     &34.1\%	&29.32\%	&0.65\%	             &7.45\%	     &0.47\%\\
mcf	               &57.56\%	     &16.48\%	&10.84\%	     &9.95\%	          &4.78\%	         &0.39\%\\
soplex	        &45.69\%	     &10.88\%	&22.76\%	&9.28\%	      &6.77\%	     &4.62\%\\
\hline
canneal	        &62.18\%	     &15.86\%	&8.9\%	     &11.57\%	 &0.91\%	 &0.58\%  \\
bodytrack	    &83.19\%	     &6.01\%	&7.66\%	&2.18\%	       &0.63\%	   &0.33\% \\
streamcluster &23.77\%	     &30.55\%	&14.38\%	&13.71\%	&17.5\%	     &0.09\%\\
\hline
DICT	            &23.86\%	&14.53\%	&28.27\%	&22.14\%	&11.06\%	     &0.14\%\\
BFS	            &3.94\%	&18.19\%	&57.42\%	&6.35\%	&5.6\%	     &8.5\%\\
setCover	    &16.26\%	&24.28\%	&27.58\%	&17.36\%	&7.5\%	     &7.02\%\\
MST	            &13.44\%	&21.28\%	&21.77\%	&25.8\%	&16.31\%	     &1.4\%\\
Graph500	    &61.48\%	&38.46\%	&0.06\%	&0\%	        &0\%	             &0\%\\
\hline
Linpack	        &22.21\%	&14.71\%	&29.18\%	&16.3\%	&9.64\%	     &7.96\% \\
\hline
NPB-CG     &0.05\%	&96.29\%	&2.66\%	&1.0\%	        &0\%	             &0\%\\
\hline
GUPS	        &95.5\%	&4.5\%	&0\%	&0\%	&0\%	     &0\% \\
\hline
\end{tabular}
\end{table}

\section{Design and Implementation}\label{sec:design-and-impl}
In this section, we first give an overview of Rainbow, and then present the technical details of hot page identification, utility-based page migration, and split TLBs. At last, we describe some other implementation issues such as cache/TLB consistency guarantees.

\subsection{Architecture Overview}

\begin{figure}[tbp]
\centering
\includegraphics[width=0.95\columnwidth]{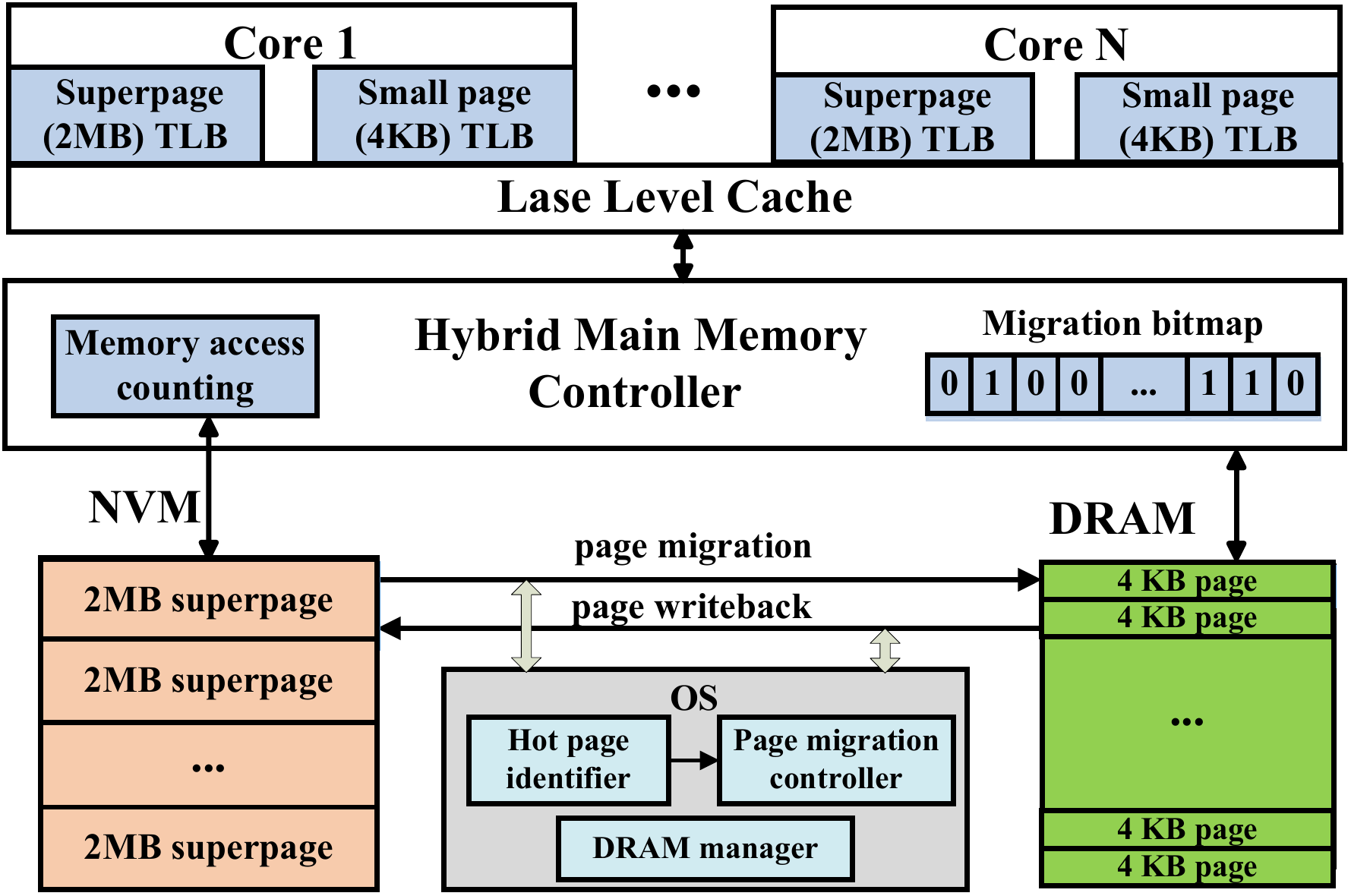}
 \vspace{-1.0ex}
\caption{Architecture of Rainbow}
\label{fig:Rainbow_arch}
 \vspace{-1.0ex}
\end{figure}

Figure~\ref{fig:Rainbow_arch} depicts the architecture of our system. Rainbow only allocates 2 MB superpages in the NVM, and use the DRAM to cache the hot NVM pages of 4 KB size. 
Correspondingly, each processor uses two split TLBs to accelerate the address translation of superpages and small pages. In the memory controller, we design a lightweight page access monitoring mechanism to identify the hot small pages in the NVM. Also, we use a migration bitmap to flag the migrated small pages on an one-bit-per-page basis.

In the operating system (OS), we develop three modules. The hot page identifier periodically reads the page access counts in the memory controller and identifies the hot small pages within the monitored superpages. The page migration controller exploits a utility-based scheme to migrate small pages when the migration benefit is expected to be larger than the migration cost. The DRAM manager is responsible for page allocation and replacement. 
We modified the buddy allocator in the OS for DRAM memory allocation. 
As the DRAM capacity is often much larger than on-chip cache, conventional LRU-based replacement policies can cause significant performance penalty when they are implemented in the software layer. Like HSCC~\cite{Liu:2017}, we use three lists to manage the DRAM memory: a free list to maintain unused pages, a clean list for unmodified pages, and a dirty list for dirty (modified) pages. Because the dirty pages should be written back to the NVM (costly), Rainbow preferentially selects free and clean pages for DRAM replacement at first. 
When the free and clean lists all become empty, Rainbow has to replace the dirty pages finally.

 \vspace{-1ex}
\subsection{Lightweight Hot Page Identification} \label{sec:Page_tracking}
Memory access monitoring at the page granularity (4 KB) is costly when the memory size becomes very large. For example, if we use only 2 bytes to record the access counts of a 4 KB page, 1 TB memory requires total $\frac{1 TB}{4 KB}\times 2B = 512$ MB storage. It is impractical to store those records in on-chip SRAM.

\begin{figure}[tbp]
\vspace{-0ex}
\centering
\includegraphics[width=\columnwidth]{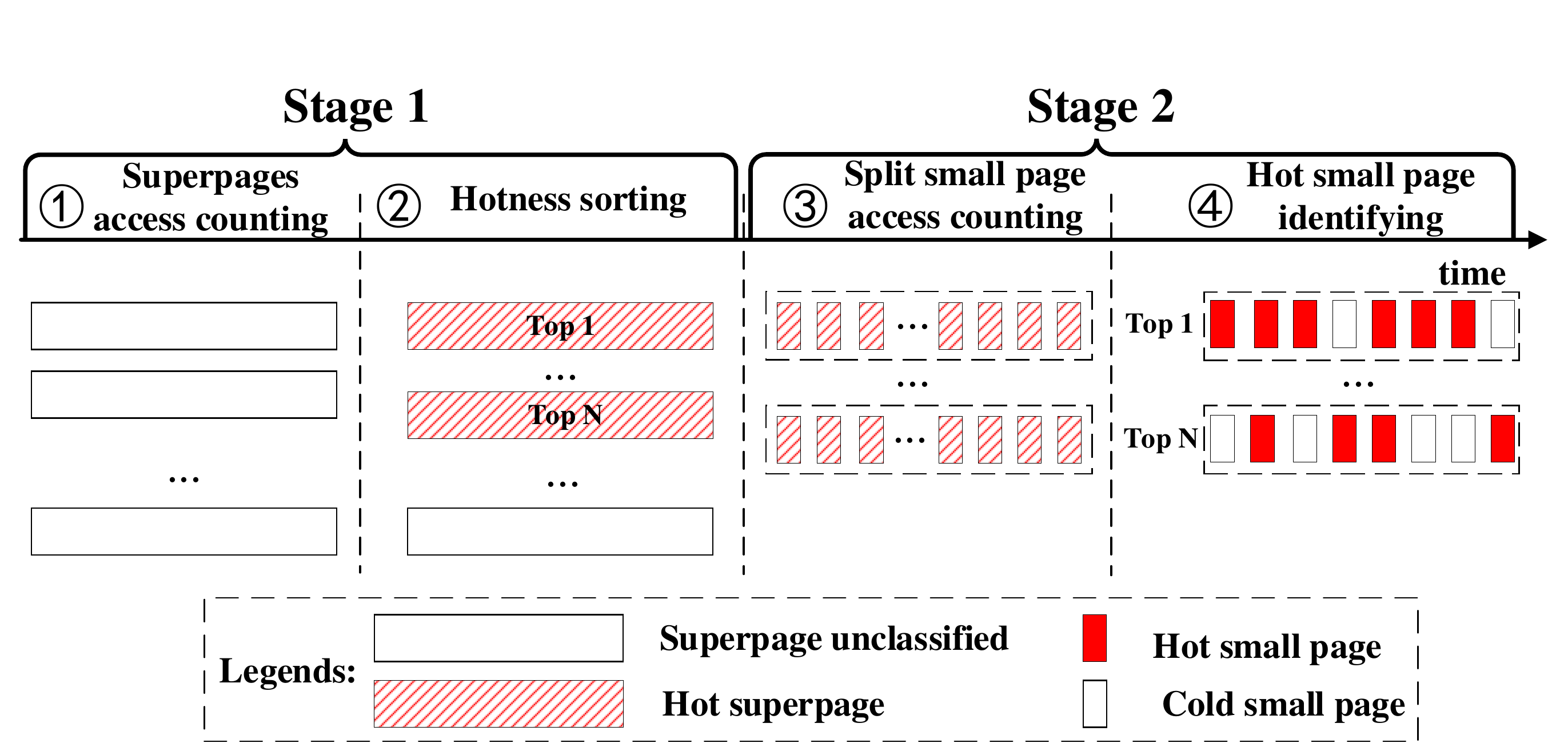}
\vspace{-3.0ex}
\caption{The two-stage memory access monitoring for small hot page identification in \textit{Rainbow}}
\label{fig:interval}
 \vspace{-2.0ex}
\end{figure}

\begin{figure}[tbp]
\centering
\includegraphics[width=0.7\columnwidth]{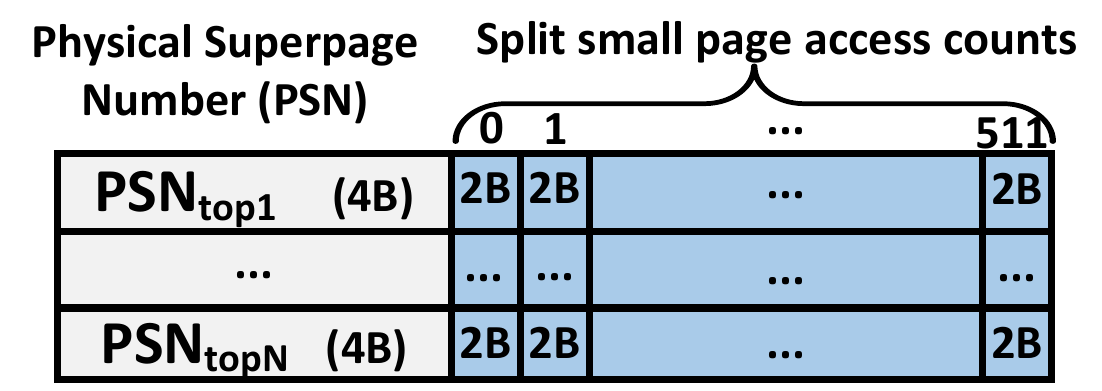}
\vspace{-0ex}
\caption{The data structure for small page access counting}
\label{fig:small_page_monitoring}
 \vspace{-2ex}
\end{figure}

We propose a two-stage memory access monitoring mechanism to mitigate the storage overhead. As shown in Figure~\ref{fig:interval}, Rainbow divides the process of memory access monitoring into two phases. In the phase 1, Rainbow first counts NVM memory accesses at the superpage granularity (\textcircled{1}). We use two bytes to store the access counts for each superpage in an interval of $10^8$ cycles. For each memory reference, the memory controller determines which superpage corresponds to the physical address and updates the access counts. We note that NVM write operations have a higher weighting of the counter value than NVM read operations.  After a given time interval for superpage access counting, Rainbow then selects the $top$ $N$ hot superpages to further monitor them at the granularity of 4 KB pages (\textcircled{2}). Even though application footprints may be very large, their working sets in a short interval is often much smaller. Thus, the superpages sorting latency is acceptable through a software approach. In the Phase 2, Rainbow monitors the small pages within those hot superpages and records their memory access counts (\textcircled{3}) in a small table.  As shown in Figure~\ref{fig:small_page_monitoring}, we need 4 bytes to record the physical superpage number, and 2 bytes to record the access counts for each small page. Note that the access counter uses 15 bits to store the data values, and 1 bit as the overflow flag. An overflow implies that the superpage is definitely hot. Thus, monitoring a hot superpage requires $4B + 512\times2B=$ 1028 bytes total storage in a fine-grained manner. At last, Rainbow classifies those split small pages into hot and cold pages via threshold based classification (\textcircled{4}). A page is determined as a hot page only when its migration benefit exceed a given threshold (see Section~\ref{sec:multi_migration}). This history-based policy avoids monitoring cold superpages at the small page granularity, and thus significantly reduce the overhead of hot page identification.

%
\subsection{Utility-based Hot Page Migration} \label{sec:multi_migration}
Page migration from NVM to DRAM can improve memory access performance and energy efficiency. However, it also incurs increased access latency of requested data. Moreover, indiscriminate page migration can result in page thrashing between DRAM and NVM when memory pressure in DRAM becomes high. We need to make a trade-off between the gained benefit and cost of page migrations. Table \ref{tab:parameters} presents the parameters for evaluating the benefits of page migrations in a time interval (10$^8$ cycles in our experiments).

When the DRAM has free pages to cache a NVM page, we should check whether the benefit of page migration is larger than the cost of page migration. We assume the migrated page will be read and written for $C_{r}$ and $C_{w}$ times in the next interval. The benefit of page migration is calculated as the total cycles saved by accessing data from DRAM against NVM.  The total cycles spent in a page migration can be deemed as a constant, as shown in Equation~\ref{equ:migration_benefit1}.
\vspace{-0.5ex}
\begin{equation}
Benefit_{mig} = (t_{nr}-t_{dr})C_{r} + (t_{nw}-t_{dw})C_{w}
         -T_{mig} 
\label{equ:migration_benefit1}
\end{equation}

\begin{table}[tbp]
 \vspace{-0ex}
\scriptsize
\centering
\caption{{\textnormal{Parameters for Evaluating Page Migrations}}}
 \vspace{-1ex}
\begin{tabular}{|c|l|}
\hline
\textbf{Notations} & \textbf{ Descriptions} \\
\hline
$C_{r}$, $C_{w}$ & total number of reads and writes on a page in a time interval\\
\hline
$t_{nr}$, $t_{nw}$ & NVM read and write latencies\\
\hline
$t_{dr}$, $t_{dw}$ & DRAM read and write latencies \\
\hline
$T_{mig}$ & cycles spent in migrating a page from NVM to DRAM  \\
\hline
$T_{writeback}$ & cycles spent in writing a dirty DRAM page to NVM\\
\hline
\end{tabular}
\label{tab:parameters}
 \vspace{-4ex}
\end{table}

When the DRAM utilization becomes high, Rainbow may need to reclaim DRAM pages for holding the newly migrated pages. Rainbow would preferentially reclaim clean pages. However, if there is no clean pages, Rainbow needs to evict dirty pages to the NVM. This results in bidirectional page migrations and less migration benefit. Assume a DRAM page $p1$ is evicted to hold a newly migrated NVM page $p2$, the total migration benefit should be offset by the cost due to page swapping, as illustrated in Equation~\ref{equ:migration_benefit2}. To mitigate the cost of page swapping, we monitor the data traffic of bidirectional page migrations, and dynamically increase the threshold of migration benefit to select hotter small pages within each superpage.
\vspace{-1.0ex}
\begin{equation}
\begin{split}
\Delta Benefit_{mig} & = (t_{nr}-t_{dr})(C_{r}^{p2}-C_{r}^{p1}) \\
   &+(t_{nw}-t_{dw})(C_{w}^{p2}-C_{w}^{p1})\\
         & -T_{mig}-T_{writeback} 
\end{split}
\label{equ:migration_benefit2}
\end{equation}


\subsection{A Small Cache for Page Migration Bitmap}

When a page is migrated to the DRAM, Rainbow sets the corresponding bit in the migration bitmap to identify whether this page has been migrated to the DRAM. For each 2 MB superpage, we need a 512-bits bitmap to record the migration flags for all 4 KB small pages. The storage overhead ($\frac{1}{4096 \times 8}$) is acceptable for a moderate-sized NVM device, and thus the migration bitmaps of all superpages can be placed in the memory controller (SRAM) for high performance. However, for large capacity memory systems, it is impractical to store all migration bitmaps in SRAM. For example, 1 TB NVM leads to 32 MB bitmap data. For better scalability, we design a small cache in the memory controller to store the migration bitmaps of recently-accessed superpages, while the whole migration bitmaps are still stored in the main memory.

\begin{figure}[tbp]
\vspace{-0ex}
\centering
\includegraphics[width=0.7\columnwidth]{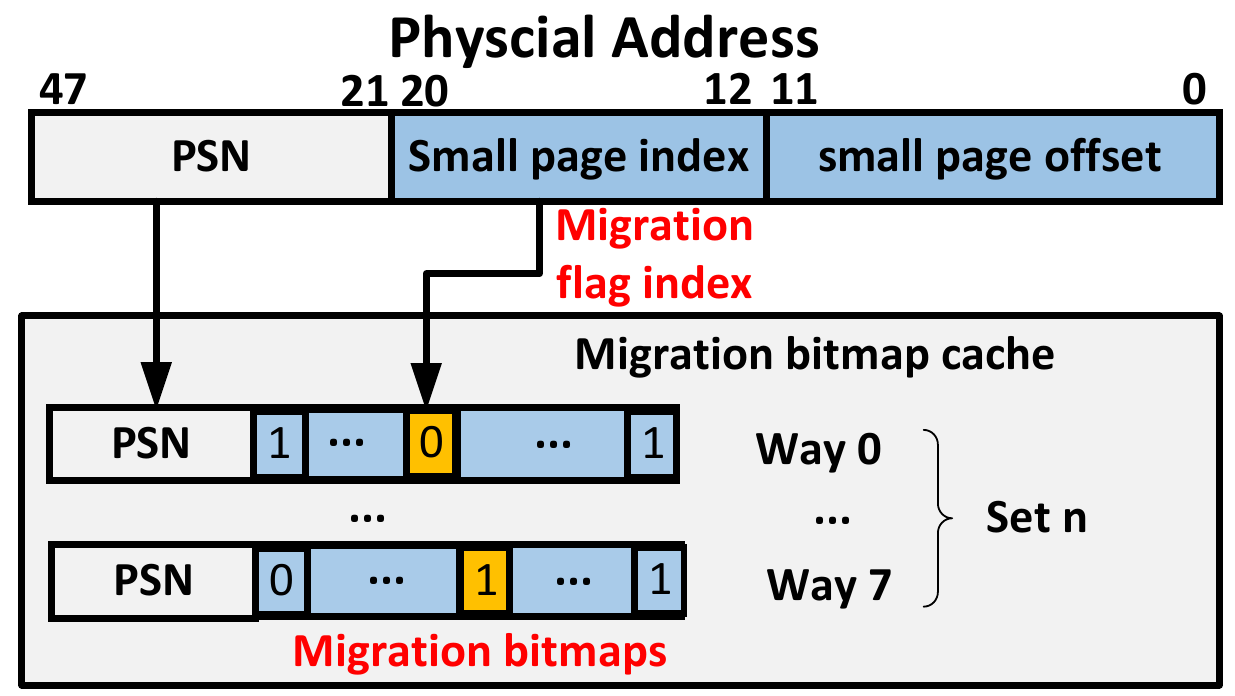}
\vspace{-0.5ex}
\caption{The migration bitmap cache in \textit{Rainbow}}
\label{fig:bitmap_cache}
 \vspace{-2ex}
\end{figure}

The migration bitmap cache is implemented as a 8-way set-associative cache, as shown in Figure~\ref{fig:bitmap_cache}. The physical superpage number (PSN) is used to index the migration bitmap of a superpage, and the middle 9 bits (12 to 20) are used to index the migration flag of a small page in the bitmap. It requires only a number of bit shifting operations to locate the migration flag. Due to space constraints in the memory controller, Rainbow only uses 4000 entries to cache the page migration flags of total 8 GB memory. Each cache entry requires 4 bytes for the PSN and 512 bits for the migration bitmap. The total storage overhead is only 272 KB SRAM. Generally, the migration bitmap cache is filled accompanying with a superpage TLB miss.
As the number of migration bitmap cache entries is much larger than the superpage TLB entries in Rainbow, the hit rate of migration bitmap cache is also higher than that of superpage TLBs. We further evaluated the timing parameters of the bitmap cache by using CACTI 3.0~\cite{CACTI-3.0}. It only leads to 9 cycles latency (similar to the L2 cache latency) before accessing the NVM, which is one order of magnitude lower than the inherent NVM access latencies.




\subsection{Split TLBs and Address Remapping}\label{sec:addressing}
Once a page has been migrated to the DRAM, Rainbow stores the destination address (DRAM page number) in the page's original place. More specifically speaking, Rainbow overwrites the beginning 8 byte data with the page's new physical address, which points to a DRAM page. Meanwhile, we set the corresponding flag in the migration bitmap. When the DRAM page is evicted, if the data is not modified (clean), we only need to write back the first 8 bytes of the page.
\begin{figure}[tbp]
\vspace{-0.0ex}
\centering
\includegraphics[width=\columnwidth]{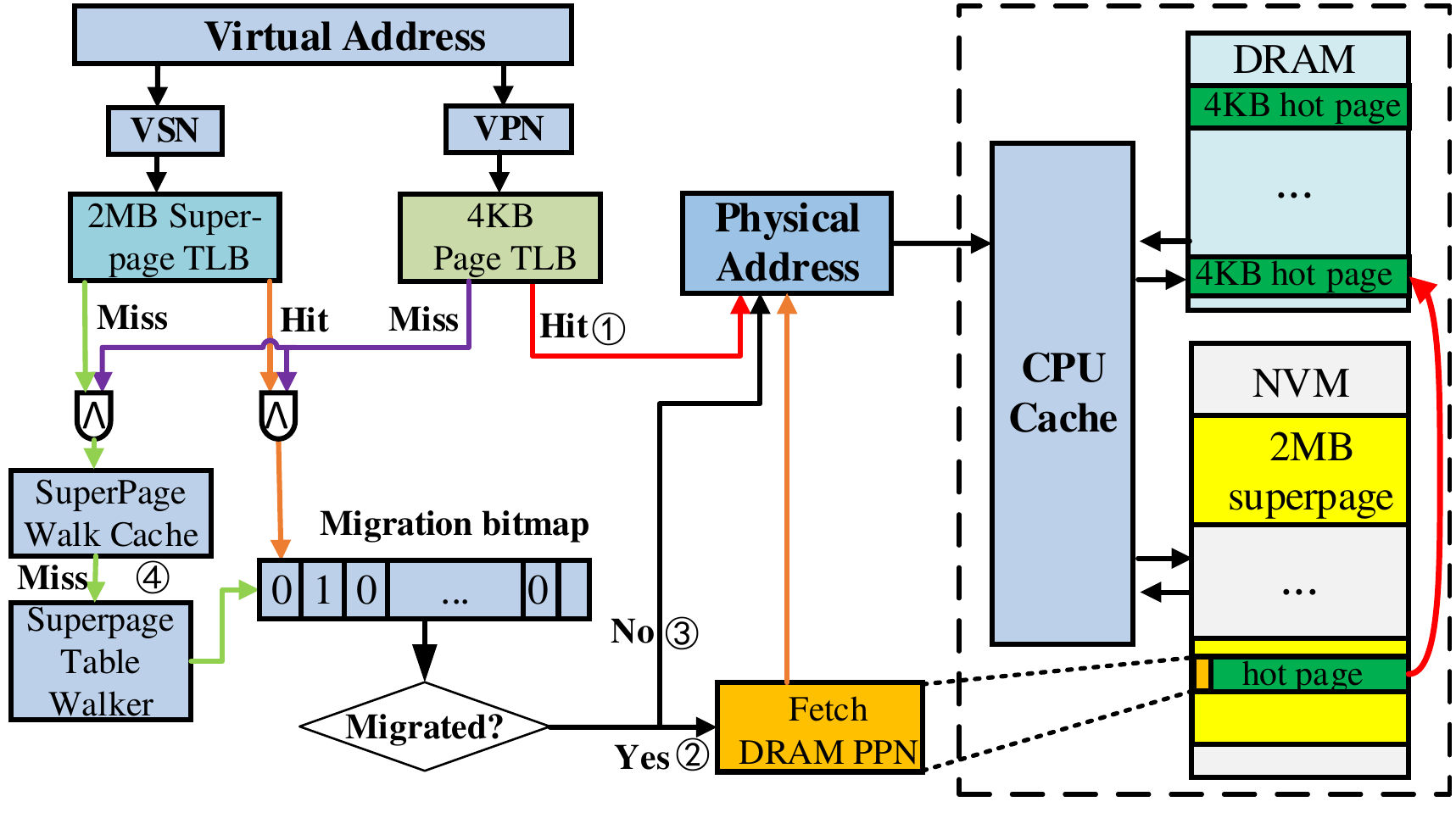}
\vspace{-3.5ex}
\caption{Four cases of memory addressing in \textit{Rainbow}}
\label{fig:addressing}
 \vspace{-3ex}
\end{figure}

Rainbow leverages split TLBs cooperatively to accelerate virtual-to-physical address translations for both DRAM and NVM. Upon a memory reference, the split TLBs are consulted in parallel, as shown in Figure~\ref{fig:addressing}. Generally, we have the following four cases: (1) 4 KB page TLB hit and superpage TLB hit; (2) 4 KB page TLB hit and superpage TLB miss; (3) 4 KB page TLB miss and superpage TLB hit; (4) 4 KB page TLB miss and superpage TLB miss.

For the first and second cases, Rainbow chooses the physical address that 
the 4 KB page TLB returns (path \textcircled{1} in Figure~\ref{fig:addressing}). These two cases imply that the accessed data has been cached in the DRAM, and the stale data in the NVM is invalid.

For the third case, Rainbow needs to check the migration bitmap with the returned physical address. If the corresponding migration bit is set, meaning that the small page within the superpage has been cached in the DRAM, Rainbow reads the first 8 bytes of this page in the NVM to obtain the destination physical address, which points to a page in the DRAM (\textcircled{2} in Figure~\ref{fig:addressing}). Otherwise (the small page is not migrated), Rainbow gets the physical address translated by the superpage TLB (\textcircled{3} in Figure~\ref{fig:addressing}). At last, Rainbow sends the translated physical address to on-chip cache or main memory (upon LLC misses) to access the requested data.

In the forth case, Rainbow performs hardware page table walking for the superpage address translation (\textcircled{4} in the Figure~\ref{fig:addressing}). When the page tables return the physical address, Rainbow also needs to check the migration bitmap, and the following operations are the same as the third case.

As illustrated in Figure~\ref{fig:addressing}, although the hot pages are migrated between DRAM and NVM, Rainbow does not need to splinter the NVM superpages and the corresponding TLBs. Any memory references to a migrated hot page are redirected to the DRAM through only one access to the superpage. This address remapping mechanism guarantees the transparency of hot page migration from the view of applications.

In the following, we analytically compare the cost of DRAM page addressing in Rainbow with the traditional page table walking mechanism~\cite{Yaniv:2016:HDC:2896377.2901456}. Upon the 4 KB page TLB miss, page table walks result in four memory references to the four-level page tables, and thus the address translation overhead is $4\times t_{dr}$. For Rainbow, we need to read the DRAM page's physical address from the corresponding superpage in NVM. Since the superpages have only three-level page tables, there are three memory references to the page tables and one memory reference for reading the DRAM page address. Assume the hit rate of superpage TLBs is $R_{hit}$, the DRAM page addressing overhead becomes  $R_{hit} \times t_{nr} +(1-R_{hit}) \times  4 t_{nr}$. Because $t_{nr}$ is almost twice as much as $t_{dr}$, we deduce that Rainbow leads to lower DRAM page addressing overhead than the page table walking mechanism when $R_{hit}>67\%$, and reduces DRAM page addressing overhead by 42.5\% when $R_{hit}=95\%$. Since the hit rate of superpage TLBs is rather high for many applications (over 99\% in our experiments), Rainbow is able to significantly reduce the overhead of DRAM page addressing. Because Rainbow can fully utilize the space of split TLBs, it essentially enables the superpage TLB to be another larger cache to the 4 KB page TLBs. 

\subsection{Data Consistency}

\textbf{Data Consistency between DRAM and NVM}. As mentioned before, the hot data has two replicas, one in DRAM and one resident in NVM. Correspondingly, a virtual address may be presented in both superpage TLBs and 
4 KB page TLBs. To guarantee the data consistency, we use a migration bitmap in the memory controller to mark the migrated hot pages.  For each memory reference on the NVM, the migration flag is first checked to make sure that Rainbow always accesses the data cached in the DRAM. 

\textbf{Cache Consistency.} Since some processors leverage write-back cache solutions, where the write operations are directed to cache and the completion is immediately confirmed to the host CPU. The dirty data blocks are written to main memory only at specified intervals or under the condition of cache evictions. This mechanism often brings higher performance but may result in inconsistency problems. Because a page may be referenced by a set of cache lines in on-chip cache, a page migration may copy the stale data to another place, allowing a portion of data inconsistent with the replica in on-chip caches. Rainbow utilizes \textit{clflush} instructions to address this problem. To be more specific, the \textit{clflush} instruction invalidates all cache lines associated with the migrated page from all levels of the processor's cache hierarchy. The invalidation is broadcast throughout the cache coherence domain. If a cache line contains modified (dirty) data at any level of the cache hierarchy, the cache line is written back to the main memory before invalidation. In this way, when a page is migrated, the corresponding dirty cache lines are written to main memory, and clean cache lines are invalidated.

\textbf{TLB Consistency.} Similar to the cache consistency issue, page migration may also cause TLB inconsistency problems~\cite{Liu:2017} because a page may be referenced by multiple cores' TLB entries. A simple solution is to adopt the TLB shootdown mechanism~\cite{Black:1989:TLB:68182.68193, Liu:2017}. That is, when a processor's TLB changes a address mapping, the same TLB entries in other cores should be invalidated. However, in Rainbow, a NVM-to-DRAM page migration do not lead to a TLB inconsistency problem. As mentioned before, our address remapping mechanism is able to logically guarantee the contiguity of superpage, and thus a hot page migration does not need to be perceived by the superpage TLB. Also, because a migrated page in DRAM is not necessarily accessed in the immediate future, the 4 KB page TLB for the DRAM page is constructed on its first access. When a DRAM page is written back to the NVM, we adopt the TLB shootdown mechanism~\cite{Black:1989:TLB:68182.68193} to invalidate all cores' 4 KB TLB entries corresponding to the DRAM page.



\section{Evaluation}\label{sec:evaluation}
\subsection{Experimental Methodology}\label{test-setup}
We implement Rainbow in an integrated simulator based on zsim~\cite{Sanchez:2013} and NVMain~\cite{NVMain}. Zsim is a fast x86-64 multi-core simulator built on Pin~\cite{Luk:2005}. We extend zsim to support many OS-level functions, such as  buddy allocator, page tables, and TLB management operations. NVMain is a cycle-accurate memory simulator that can model both DRAM and NVM in detail. In our experiment, NVMain is used to simulate the hybrid main memory composed of DRAM and NVM, each with an individual memory controller. 

\textbf{Configuration.} The platform and the detailed configuration in our experiments are depicted in Table~\ref{tab:testbed}. PCM is chosen as the storage medium of main memory as it is a widely studied NVM. Timing and energy parameters of PCM are referred to papers~\cite{Lee:2009:APC:1555754.1555758,Liu:2017}. We also model the latencies of clflush, TLB shootdown, and data move in details according to the timing parameters of CPU and DRAM/NVM. 

\begin{table}[tbp]
\vspace{-0.5ex}
\scriptsize
\centering
\caption{{\textnormal{System Configuration in Rainbow}}}
 \vspace{-2ex}
\setlength{\tabcolsep}{0.9mm}{
\begin{tabular}{|l|l|}
\hline
CPU & 8 cores, 3.2 GHz, out-of-order \\
\hline
L1 TLB & \tabincell{l}{32 Data TLB entries for 2 MB superpages, and 32 Data TLB entries\\ for 4 KB small pages per core, 4-way, 1-cycle latency} \\
\hline
L2 TLB & \tabincell{l}{512 unified TLB entries for 2 MB superpages, and 512 unified TLB\\ entries for 4 KB small pages, 8-way, 8-cycle latency} \\
\hline
L1 Cache & \tabincell{l}{private 64 KB per core, 4-way, split D/I,  3-cycle latency} \\
\hline
L2 Cache & \tabincell{l}{private 256 KB per core, 8-way, 10-cycle latency} \\
\hline
L3 Cache & \tabincell{l}{shared 8 MB, 16-way, 34-cycle latency} \\
\hline
Bitmap cache & \tabincell{l}{272 KB, 8-way, 9-cycle latency} \\
\hline
\tabincell{c}{DRAM} & \tabincell{l}{4 GB: 1 channel, 4 rank, 32 banks, 32768 rows, 64 cols,\\
                 Bandwidth: 10.7 GB/Sec, FR-FCFS request scheduling,  \\
                Timing (tCAS-tRCD-tRP-tRAS) : 7-7-7-18 (cycles),\\ 13.5 ns read latency, 28.5 ns write latency }\\

\hline
\tabincell{c}{PCM} & \tabincell{l}{32 GB: 4 channels, 8 ranks, 8 banks per rank,  65536 rows, 32 cols,\\
                   Bandwidth: 10.7 GB/Sec, FR-FCFS request scheduling, \\
                  Timing (tCAS-tRCD-tRP-tRAS): 9-37-100-53 (cycles), \\19.5ns read latency, 171 ns write latency \\
                   }\\
\hline
\multicolumn{2}{|c|}{\textbf{Power/Energy consumption}}\\
\hline
DRAM & \tabincell{l}{Voltage: 1.5V,   Standby: 77 mA, Refresh: 160 mA, Precharge: 37 mA;\\ Read and write on row buffer hit: 120 and 125 mA; \\
Read and write on row buffer miss: 237 and 242 mA
         }\\
\hline
PCM & \tabincell{l}{Read/write on row buffer hit: 1.616 pJ/bit; \\
Read and write on row buffer miss: 81.2 pJ/bit and 1684.8 pJ/bit}\\
\hline
\end{tabular}}
\label{tab:testbed}
 \vspace{-1.5ex}
\end{table}

\textbf{Alternative policies.}
We compare Rainbow with several alternative page migration policies for hybrid memories as follows.
(1) \textbf{Flat-static}: 4 GB DRAM and 32 GB NVM are organized in a flat address space~\cite{Liu:2017}, and are managed in 4 KB small pages. Data is evenly distributed in DRAM and NVM according to the ratio of DRAM to NVM capacity (1:8). There is no page migration between NVM and DRAM. We use this system as a baseline for comparison.
(2) \textbf{HSCC-4KB-mig}: HSCC is a state-of-the-art hybrid memory system that supports utility-based page migration at the granularity of 4 KB page~\cite{Liu:2017}. The major difference between Rainbow and HSCC is the support of superpages. This comparison is made to evaluate the effectiveness of using superpages in hybrid memory systems.  
(3)\textbf{ HSCC-2MB-mig}: we modify HSCC to support superpages and memory migration at 2 MB superpages granularity. This comparison is made to evaluate the performance and energy penalties of superpage migrations.
(4) \textbf{DRAM-only}: this is a system with only 32 GB DRAM and supports only 2 MB superpages. We use it as the applications' performance upper bound because they can benefit from superpages without any page migrations.



\textbf{Benchmarks.}
We evaluate a number of workloads with different memory access patterns from SPEC CPU 2006~\cite{Spec_CPU2006}, Parsec~\cite{Parsec}, Problem Based Benchmarks Suite (PBBS)~\cite{PBBS}, Graph500\cite{Graph500}, Linpack\cite{Linpack}, NPB-CG~\cite{CG}, and GUPS~\cite{GUPS}, as listed in Table~\ref{tab:workload}. Detailed memory access patterns of these applications are shown in Table~\ref{tab:hotpage}. In addition, we evaluate three multi-programmed workloads, as shown in Table~\ref{tab:workload}.


\begin{table}[tbp]
 \vspace{-0.0ex}
\centering
\caption{{\textnormal{Workloads for Evaluation}}}
 \vspace{-1ex}
 \setlength{\tabcolsep}{1mm}{
\begin{tabular}{|c|c|}
\hline
\textbf{Workloads} & \textbf{Applications} \\
\hline
SPEC CPU2006 & cactusADM, mcf, soplex \\
\hline
Parsec & canneal, bodytrack, streamcluster\\
\hline
PBBS & \tabincell{l} {DICT, BFS, setCover, MST} \\
\hline
{Large footprints} & {Graph500,  Linpack, NPB-CG, GUPS}  \\
\hline
{mix1} & {cactusADM+soplex+setCover+MST}\\
{mix2} &	{setCover+BFS+DICT+mcf}\\
{mix3} & {canneal+DICT+MST+soplex}\\
\hline
\end{tabular}}
\label{tab:workload}
 \vspace{-3ex}
\end{table}
\subsection{Address Translation Overhead}\label{sec:results}


\begin{figure}[tbp]
 \vspace{-0ex}
\centering
\includegraphics[width=1\columnwidth]{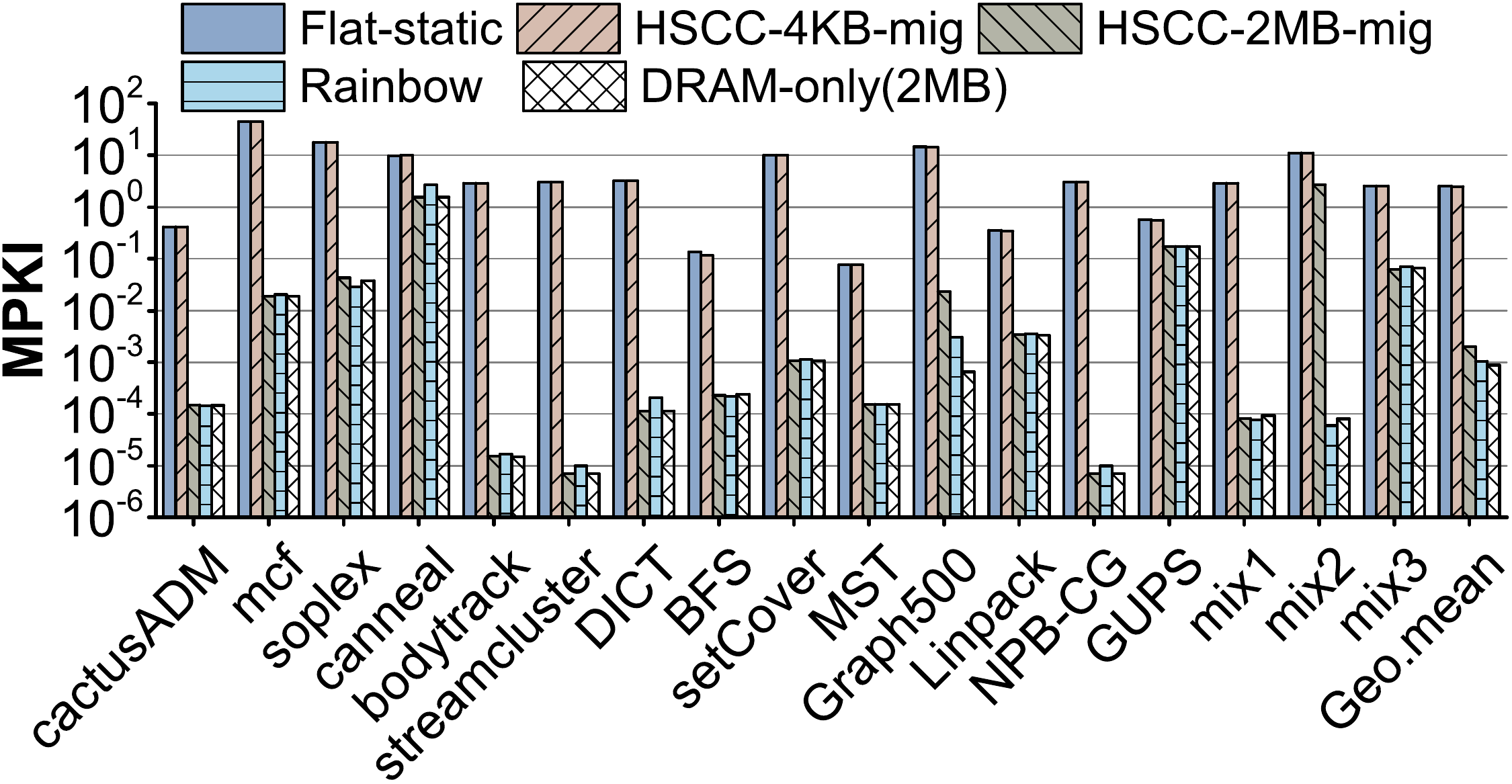}
\vspace{-3ex}
\caption{MPKI of applications}
\label{fig:TLB_perf}
 \vspace{-2ex}
\end{figure}
Figure~\ref{fig:TLB_perf} shows superpages significantly reduce TLB misses per kilo instructions (MPKI) by several orders of magnitude on average. Although Rainbow supports different page sizes, it shows almost similar TLB performance with HSCC-2MB-mig and DRAM-only (2 MB). The reason is that Rainbow logically uses the superpage TLBs as a larger next-level cache to the 4 KB page TLBs.  For applications with small memory footprints, such as bodytrack, and streamcluster, the MPKI is significantly reduced because of the wider TLB coverage (1 GB) offered by the superpages. As shown in Table~\ref{tab:hotpage},  
GUPS and \textit{canneal} are memory intensive benchmarks and show very large working sets in a short sampling interval. As a result, GUPS and \textit{canneal} show relatively high MPKI even using superpages.
Mix2 shows both a large working set and large memory footprint, leading to a large amount of page swapping between DRAM and PCM in HSCC-2MB-mig. Thus, HSCC-2MB-mig incurs a lot of TLB shootdown operations, causing a relatively high MPKI. In contrast, Rainbow does not cause TLB shootdown when migrating hot small pages within superpages to DRAM.

\begin{figure}[tbp]
 \vspace{-0ex}
\centering
\includegraphics[width=1\columnwidth]{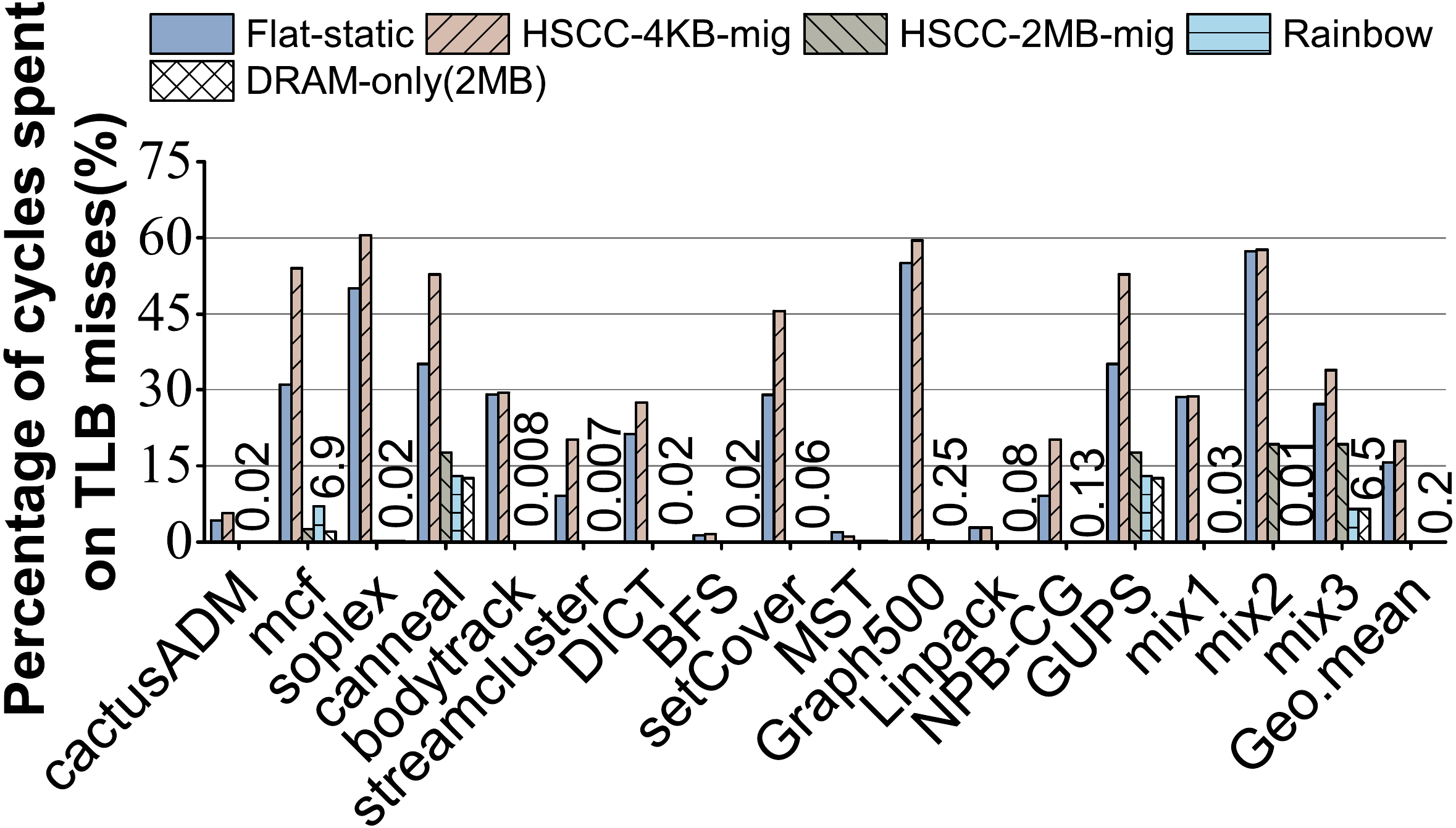}
\vspace{-3ex}
\caption{Percent of total cycles spent on servicing TLB misses. The very small values show TLB miss overheads in Rainbow.}
\label{fig:addr_trans_overhead}
 \vspace{-2ex}
\end{figure}

Figure~\ref{fig:addr_trans_overhead} shows the percent of execution cycles spent on servicing TLB misses for different applications. When the memory are managed in 4 KB small pages, the TLB miss overhead even approximates to 60\% of total execution cycles for soplex, Graph500 and mix2. For mcf, canneal, GUPS, and mix3, since their working sets approximate or exceed the superpage TLB coverage, they cause relatively high address translation overheads even using superpages.  Overall, superpages are able to significantly reduce TLB miss overhead by 99.8\% on average.

\begin{figure}[tbp]
 \vspace{-0ex}
\centering
\includegraphics[width=1\columnwidth]{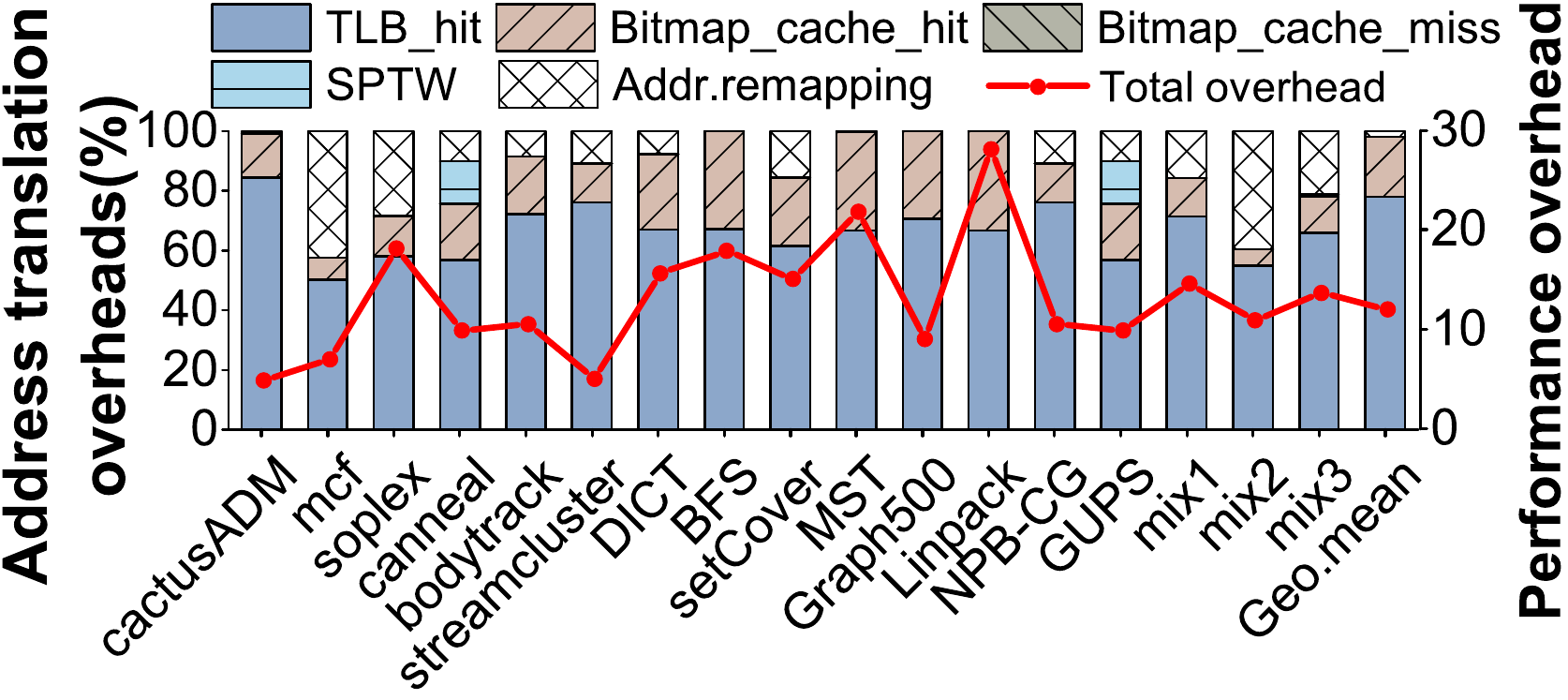}
\vspace{-3ex}
\caption{The breakdown of detailed address translation overheads in Rainbow.}
\label{fig:addr_trans_overhead_rainbow}
 \vspace{-2.5ex}
\end{figure}
We further study the detailed address translation overheads in Rainbow. Figure~\ref{fig:addr_trans_overhead_rainbow} shows the breakdown of execution cycles spent on split TLB hits, bitmap cache hits/misses, superpage table walks (SPTWs), and address remapping. The overall address translation only cause 12\% performance overhead on average. Since split TLBs are on the critical path of address translation, they introduce 78.5\% of total address translation overhead although the TLB latency is very short. The bitmap cache costs near 20\% of total address translation overhead because it should be consulted for each access to the NVM. To address DRAM pages when the corresponding 4 KB page TLB misses, the address remapping mechanism leads to 1.4\% of total address translation overhead on average. The bitmap cache miss can result in relatively higher latency, however, we observed trivial bitmap cache misses even for applications with very large footprints, such as Graph500, NPB-CG, and Linpack. As the superpage hit rate even exceeds 99.9\% on average, the average cost on superpage table walks is as low as 0.1\%. We only observe remarkable cost on SPTWs for \textit{canneal} and GUPS, because their large working sets lead to relatively lower superpage hit rate.

\subsection{Application Performance}\label{sec:IPC}
\begin{figure}[tbp]
 \vspace{-0ex}
	\centering
	\includegraphics[width=1\columnwidth]{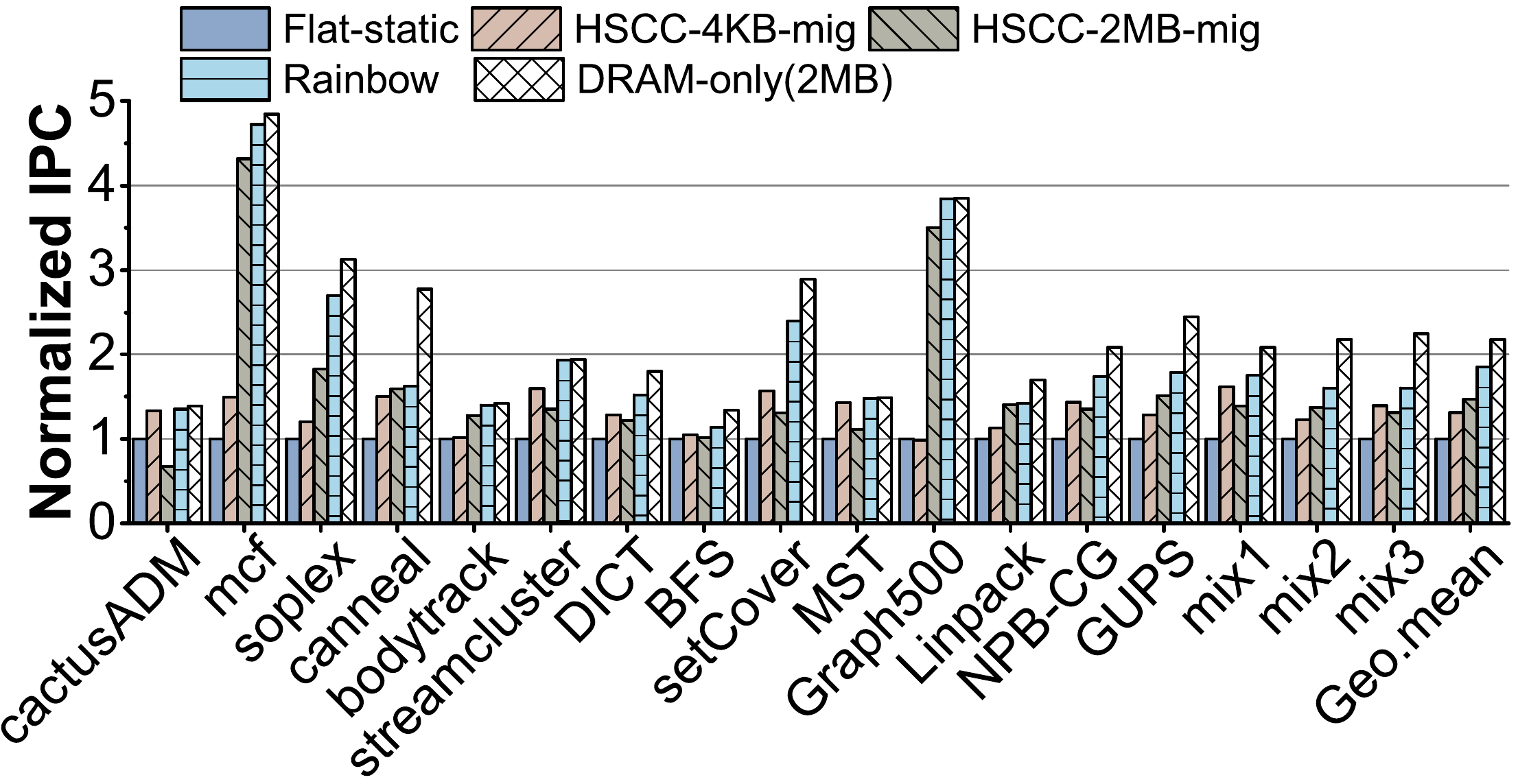}
    \vspace{-3ex}
	\caption{Normalized IPC relative to the Flat-static system}
	\label{fig:ipc}
	\vspace{-2.5ex}
\end{figure}

Figure~\ref{fig:ipc} shows the \textit{instructions per cycle} (IPC) of each application normalized to the baseline system (Flat-static). Rainbow achieves 72.7\%, 22.8\%, and 17.3\% performance improvement on average compared to Flat-static, HSCC-4KB-mig, and HSCC-2MB-mig, respectively. The performance difference between Rainbow and the upper bound (DRAM-only) is only 14.0\% on average.


Compared to HSCC-4KB-mig, Rainbow can significant improve the IPC of \textit{mcf}, \textit{soplex}, and \textit{Graph500} by 2.1X, 1.2X, and 2.9X, respectively.  For \text{mcf}, 
since superpages reduce the MPKI by approximate 99\%, they deliver 2.1X performance improvement compared to HSCC-4KB-mig. This suggests that superpages can significantly reduce the address translation overheads for memory-intensive applications. \textit{Soplex}, \textit{SetCover}, GUPS, and \textit{Graph500} all show rather poor data locality. However, they also show significant performance improvement against the systems without superpage support. This implies that applications with poor data locality can also benefit from superpages because of the improved TLB coverage. 

We also find that HSCC-2MB-mig may result in lower application performance than HSCC-4KB-mig, such as \textit{cactusADM, streamcluster, DICT, setCover, NPB-CG} and \textit{MST}. This implies that page migrations at the superpage granularity are extremely costly. The benefit of using superpages is significantly offset by the cost of superpage migrations. In contrast, Rainbow explores the advantages of both superpages and lightweight page migrations, and thus achieve much better application performance. For the DRAM-only system with 2 MB superpages supported, it shows the best performance against other policies because it takes full advantage of superpages without any memory migrations. We also note that it is not a completely fair comparison, since DRAM-only uses more DRAM.


\textbf{Insight}. \textit{(1) For applications with intensive memory accesses or poor data locality, Rainbow can significantly improve application performance by up to 2.9X. (2) For other applications, the cost of superpage migrations can offset the advantages of superpages. Using the proposed lightweight page migration scheme without splintering superpages, Rainbow can still improve application performance by 37.4\%. }

\subsection{Page Migration Traffic}
\begin{figure}[tbp]
    \vspace{-0ex}
	\centering
	\includegraphics[width=1\columnwidth]{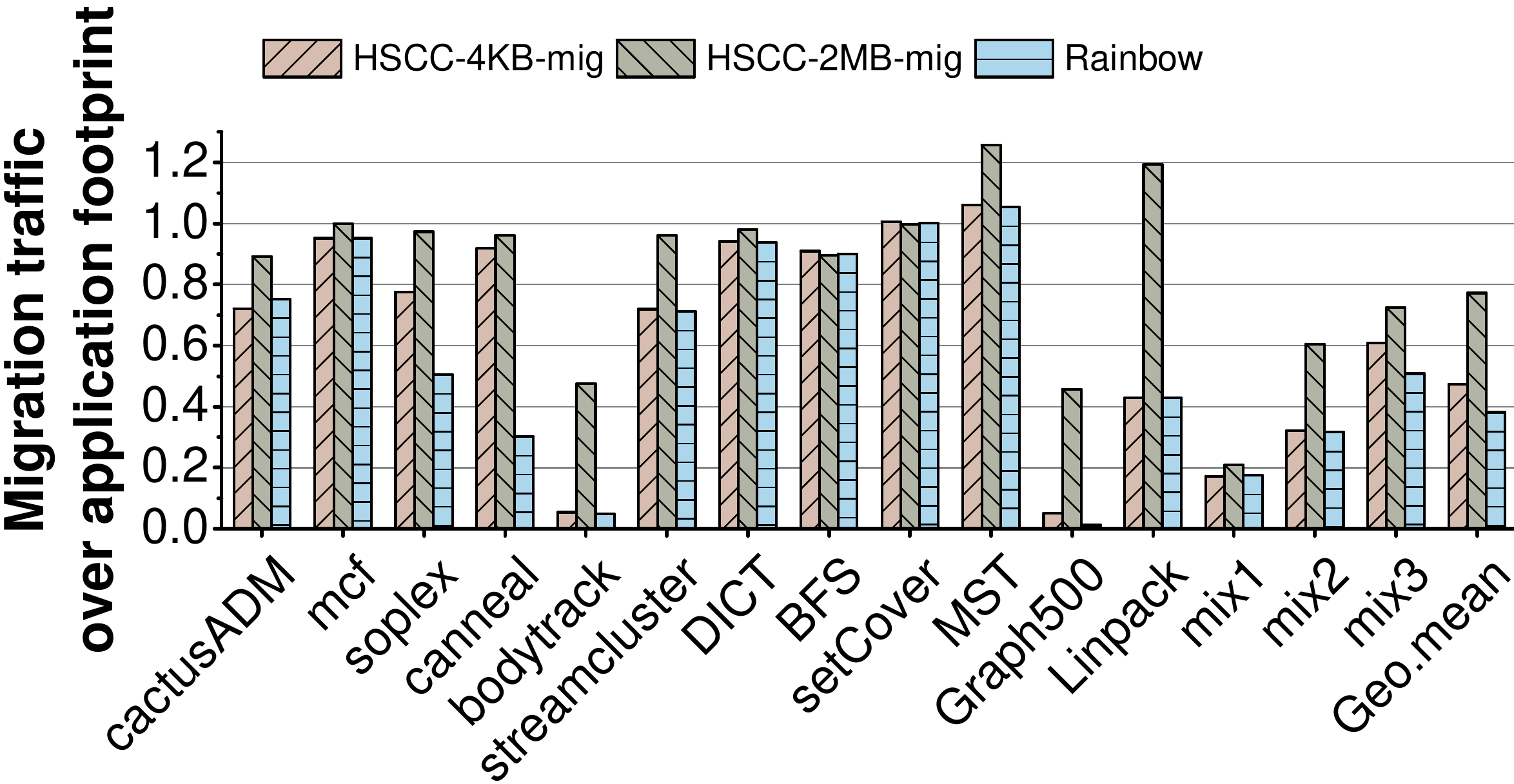}
    \vspace{-3.0ex}
	\caption{Normalized page migration traffic relative to the applications' total memory footprints}
	\label{fig:migrate_data}
	\vspace{-1.0ex}
\end{figure}

Figure~\ref{fig:migrate_data} shows the ratio of migration traffic to total memory footprint for each application.  Generally, HSCC-2MB-mig shows larger migration traffic than other migration polices because of the large granularity of page migrations. As a result, superpage migrations lead to wasted memory bandwidth on copying the cold data within superpages. Rainbow can reduce page migration traffic by 50\% on average compared to HSCC-2MB-mig. For memory-intensive applications, such as soplex, canneal, and Graph500, HSCC-4KB-mig shows more migration traffic than Rainbow because page access counting scheme in HSCC is implemented in TLB and does not filter the memory references in on-chip caches, and thus more pages are migrated to the DRAM. For MST, GUPS and Linpack, because their memory footprints are larger than the capacity of DRAM (4 GB), HSCC-2MB-mig leads to a large amount of page swapping between DRAM and PCM. Thus, the migration traffic is even larger than their total memory footprints. In contrast, Rainbow only select the hot small pages for migration, and thus significantly mitigate the frequent page swapping. We also observe that page migrations only consume 1.35\% of total memory bandwidth at most. Thus, page migrations lead to trivial memory bandwidth contention with these applications.

\subsection{Energy Consumption}
\begin{figure}[tbp]
	\centering
	\includegraphics[width=1\columnwidth]{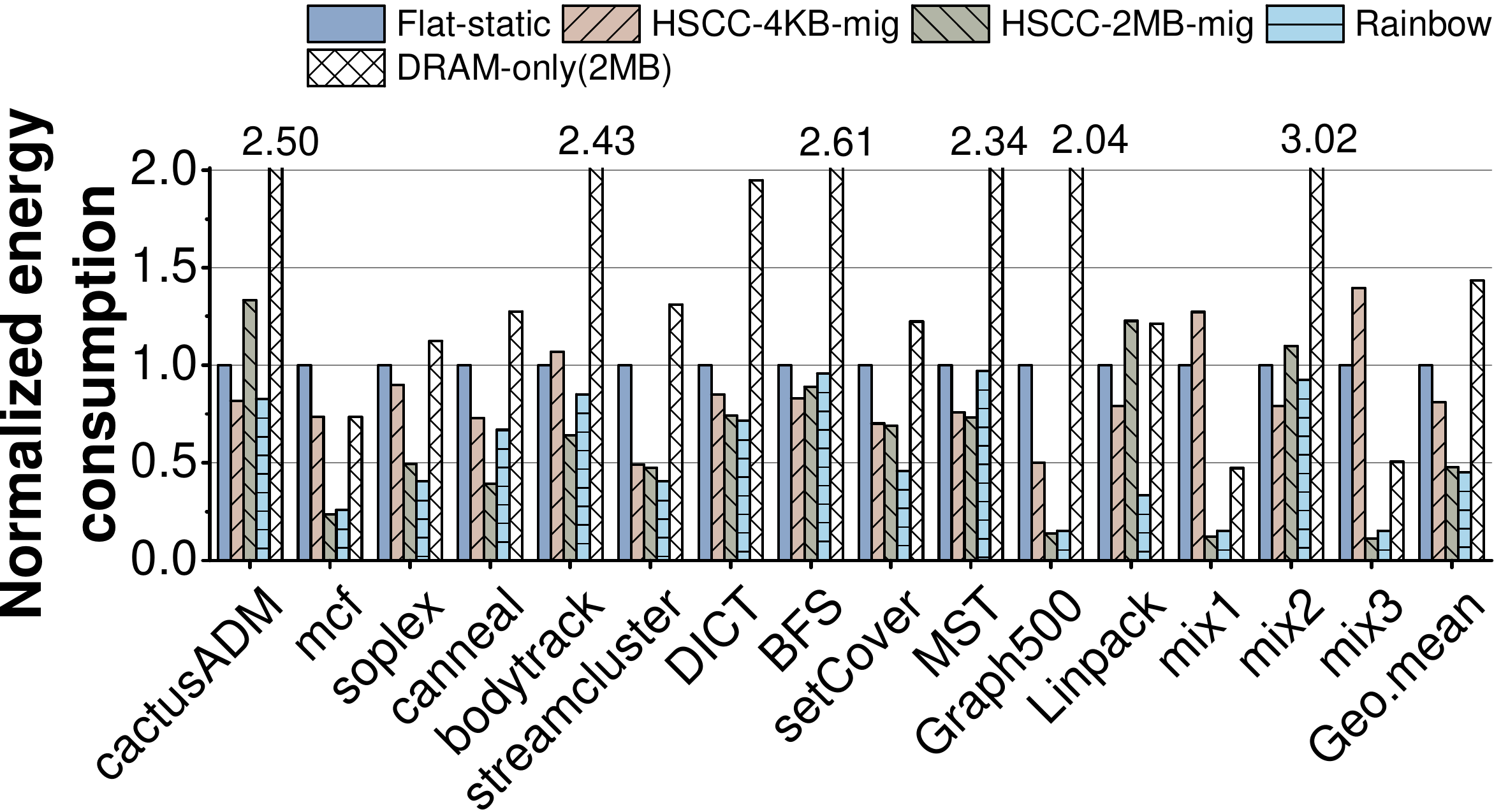}
    \vspace{-3.0ex}
	\caption{Normalized energy consumption relative to the baseline system}
	\label{fig:energy}
	\vspace{-3ex}
\end{figure}
DRAM consumes a large amount of energy due to periodical refreshing, while NVM leads to near-zero static energy consumption. To evaluate energy efficiency of Rainbow, we compare energy consumption of those migration schemes using Flat-static as a baseline, as shown in Figure~\ref{fig:energy}. Generally, the DRAM-only system shows much more energy consumption than the hybrid memory systems on average. Rainbow is able to reduce energy consumption by 45.1\% and 68.5\% on average compared to the Flat-static and DRAM-only (2 MB), respectively. Although Flat-static does not introduce additional energy consumption due to page migrations, it causes more energy consumption than HSCC and Rainbow. The reason is that a large amount of memory references are distributed on the PCM, resulting in higher active energy consumption of PCM.  Because write operations on PCM consumes 20 times more energy per bit than on DRAM~\cite{Lee:2009:APC:1555754.1555758}. This phenomenon is more clear for \text{mcf}, which shows that the misuse of hybrid memories even causes higher energy consumption compared to the DRAM-only system. In contrast, Rainbow and HSCC migrate hot pages to the DRAM, which services a large portion of memory accesses with higher energy efficiency. 

\subsection{Sensitivity Studies}
\begin{figure}
  \centering
  \vspace{-0ex}
   \subfigure{
      \label{fig:subfig:sensitive_interval_migrate_data} 
      \includegraphics[width=1.56in]{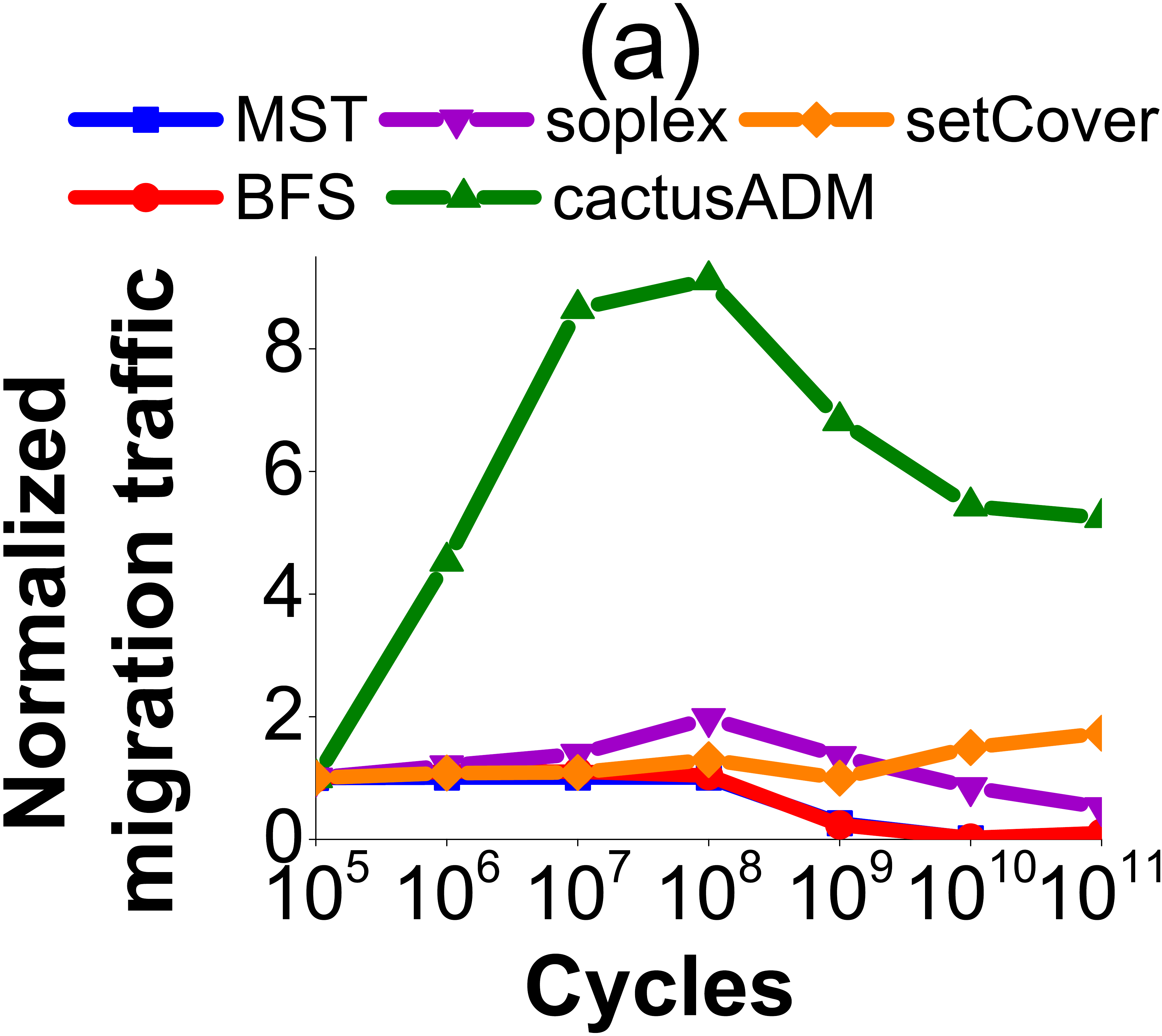}
    }
  \subfigure{
    \label{fig:subfig:sensitive_interval_ipc} 
    \includegraphics[width=1.56in]{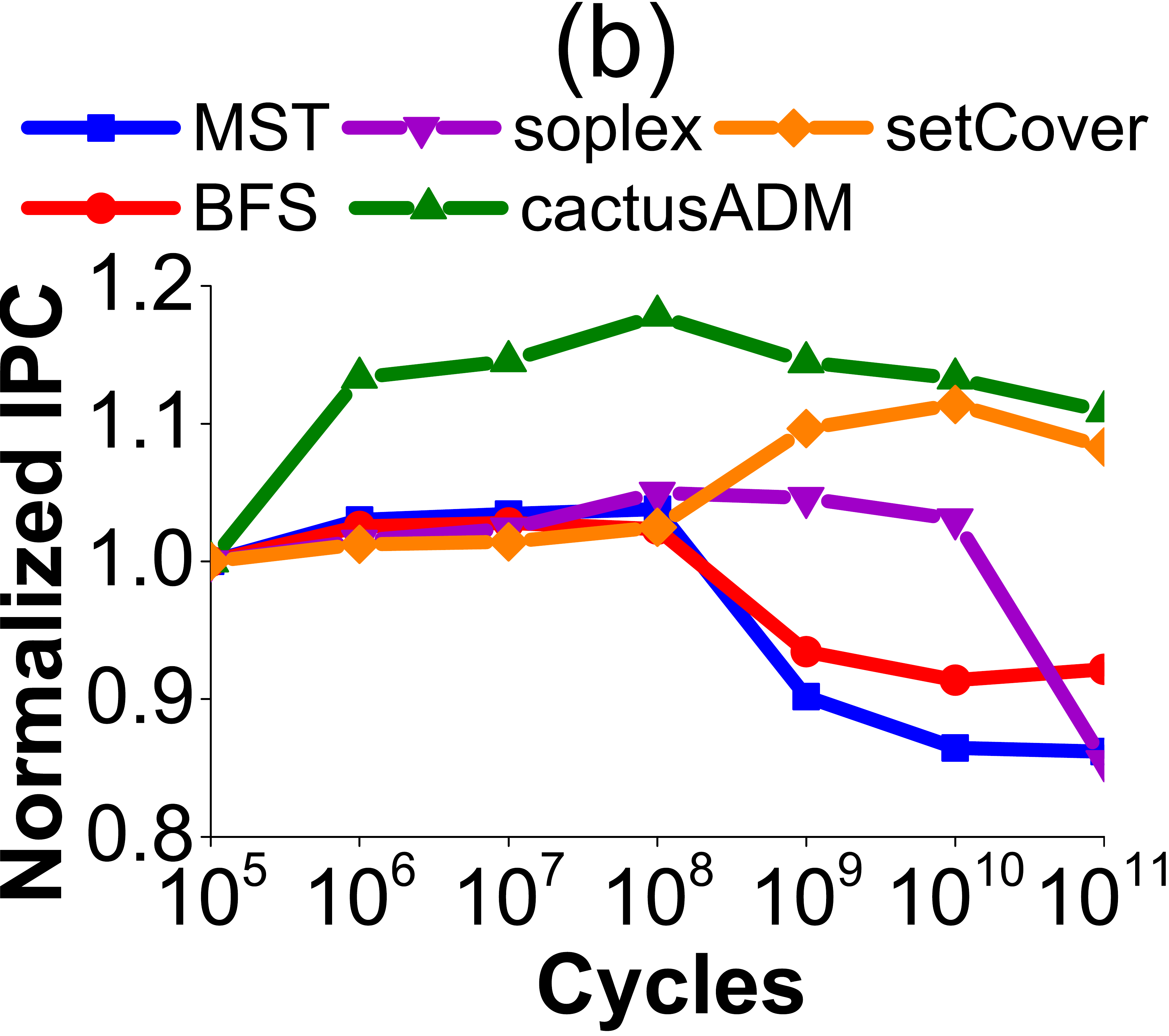}
  }

  \vspace{-2.5ex}
  \caption{ Migration traffic and IPC vary with the sampling intervals in Rainbow}
  \vspace{-2ex}
\end{figure}

To study how the application performance in Rainbow is sensitive to the time interval for hot page monitoring, we run selected applications with different sampling intervals. Note that we increase the interval and the number of monitored top $N$ hot superpages by the same factor (10). Figure~\ref{fig:subfig:sensitive_interval_migrate_data} and  Figure~\ref{fig:subfig:sensitive_interval_ipc} show how the normalized  migration traffic and application IPC are sensitive to the sampling interval, respectively. All experimental results are normalized to the setting of $10^5$ cycles. Generally, a longer sampling interval usually cause less software overhead for hot page identification. However, we find that less hot pages are migrated to the DRAM when the sampling interval exceeds $10^8$ cycles. Correspondingly, the applications' IPC also decreases with the growth of sampling interval.  As a result, we set the sampling interval as $10^8$ cycles in Rainbow for better performance.


\begin{figure}
  \centering
  \vspace{-0ex}
  \subfigure{
    \label{fig:subfig:sensitive_topn_migrate_data} 
    \includegraphics[width=1.56in]{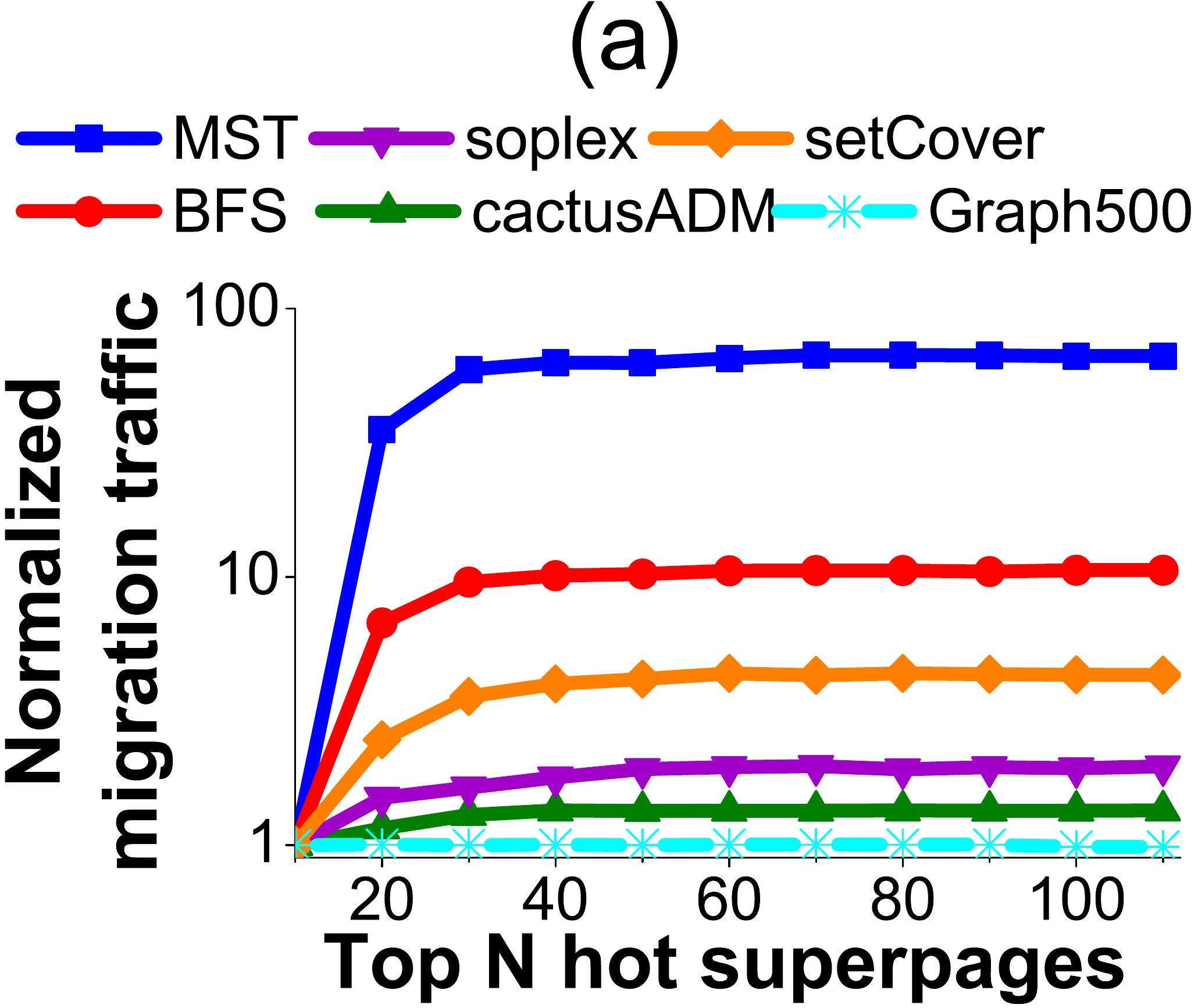}
  }
  \subfigure{
    \label{fig:subfig:sensitive_topn_ipc} 
    \includegraphics[width=1.56in]{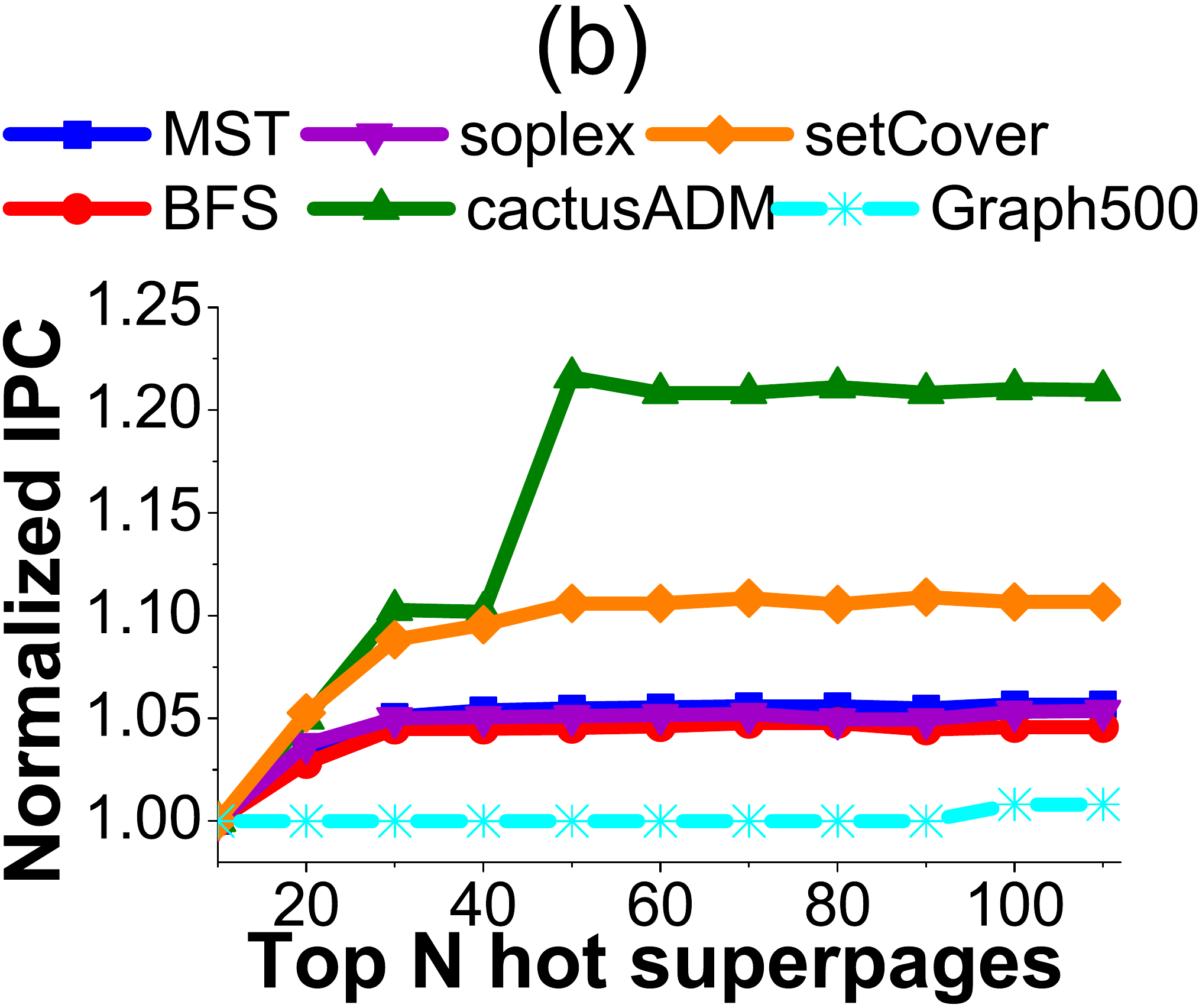}
  }
  \vspace{-1.0ex}
  \caption{Migration traffic and IPC vary with the number of top \textit{N} hot superpages in Rainbow}
  \vspace{-1ex}
\end{figure}

To evaluate how the number of selected top \textit{N} hot superpages can affect the page migration traffic and application performance in a given time interval (10$^8$ cycles), we run some memory-intensive applications with different settings of \textit{N}. Figure~\ref{fig:subfig:sensitive_topn_migrate_data} shows that there is trivial growth of migration traffic for those applications when the number of selected hot superpages exceeds 50. Figure~\ref{fig:subfig:sensitive_topn_ipc} also shows that those applications' IPC become stable when the value of $N$ is larger than 50. This suggests that the majority of hot small pages of applications are distributed on only a few hot superpages. As a result, we prudently set $N$ to be 100 in Rainbow. We argue that the top 100 hot superpages are enough for hot small pages identification, because many applications' working sets are much less than 200 MB in each sampling interval.

We have also studied the sensitivity of other settings in Rainbow. Due to space limitation, we only describe the results here. The first one is the threshold for hot page identification. We find that less hot pages are migrated to DRAM when the threshold increases. Correspondingly, the applications' IPC also become lower. We have also studied the impact of different NVM access latencies on the effectiveness of page migration. When the NVM read/write latencies increase linearly, a little more pages are migrated to DRAM because the migration benefit increases according to Equation~\ref{equ:migration_benefit1} and Equation~\ref{equ:migration_benefit2}. However, the applications' performance degrades because a large portion of cold data on the NVM introduce higher accumulative access delay.


%

%
%
%
%

\subsection{Storage and Runtime Overhead}\label{sec:overheads}

\begin{table}[tbp]
 \vspace{-0ex}
	\scriptsize
	\centering
	\caption{{\textnormal{Storage Overhead of Rainbow with 1 TB PCM}}}
	\vspace{-1.5ex}
	\begin{tabular}{|c|c|}
		\hline
		\textbf{Data Structure} & \textbf{Storage overhead} \\
		\hline
		\tabincell{l} {Migration bitmap (1 bit per 4 KB-sized page )} &
        \tabincell{l} {\textcolor{black}{272 KB} SRAM }\\
		\hline
		 \tabincell{l} {Access counters for superpages in PCM (2 Byte per 2 MB )}&   1 MB SRAM \\
		\hline
		 \tabincell{l} {Physical Superpage number (PSN) of top $N$  hot superpages \\(4 Byte per superpage )} & $4N$ Bytes SRAM\\
		\hline
		 \tabincell{l} {Access counters for split small pages in the top $ N$ hot \\ superpages  ($2B \times 512=1 KB$ per hot superpage) }& { $N$ KB} SRAM \\
		\hline
		\hline
		Total storage overhead (if $N=100$) &  \tabincell{l}{ \textcolor{black}{1.372 MB SRAM} }\\
		\hline
	\end{tabular}
	\label{tab:storageoverhead}
	\vspace{-2.5ex}
\end{table}

We analyze the storage overhead of Rainbow in a hybrid memory system comprising of 1 TB  PCM.  
\textcolor{black}{The storage overheads mainly come from migration bitmaps and superpage access counters.} We list all costs in Table~\ref{tab:storageoverhead}.
\textcolor{black}{For the migration bitmaps, 1 TB PCM needs total $\frac{1 TB}{4 KB\times 8} = 32 MB$ storage to store all superpages' migration bitmaps. We put the whole bitmaps in the main memory, and cache only a portion of recently-accessed ones} \textcolor{black} {(272 KB)} in the memory controller.
We use 2 bytes to record both superpages and small pages' access counts. There are total 512K superpages for 1 TB PCM, and thus consumes 1 MB SRAM for superpage access counters.  As described in Section~\ref{sec:motivation}, although many applications may have a very large memory footprint during execution, they show relatively small working sets in the sampling time interval ($10^8$ cycles).  As a result,
we only select the top 100 hot superpages for fine-grained page access counting at the second stage, and thus only requires $1.004*100=100.4$ KB additional storage. Overall, Rainbow only causes 1.372 MB SRAM storage overhead for a big memory system with 1 TB PCM, and the hardware die area overhead modeled by CACTI~\cite{CACTI-3.0} is only 7\%. 

\begin{figure}[tbp]
	\centering
	\includegraphics[width=1\columnwidth]{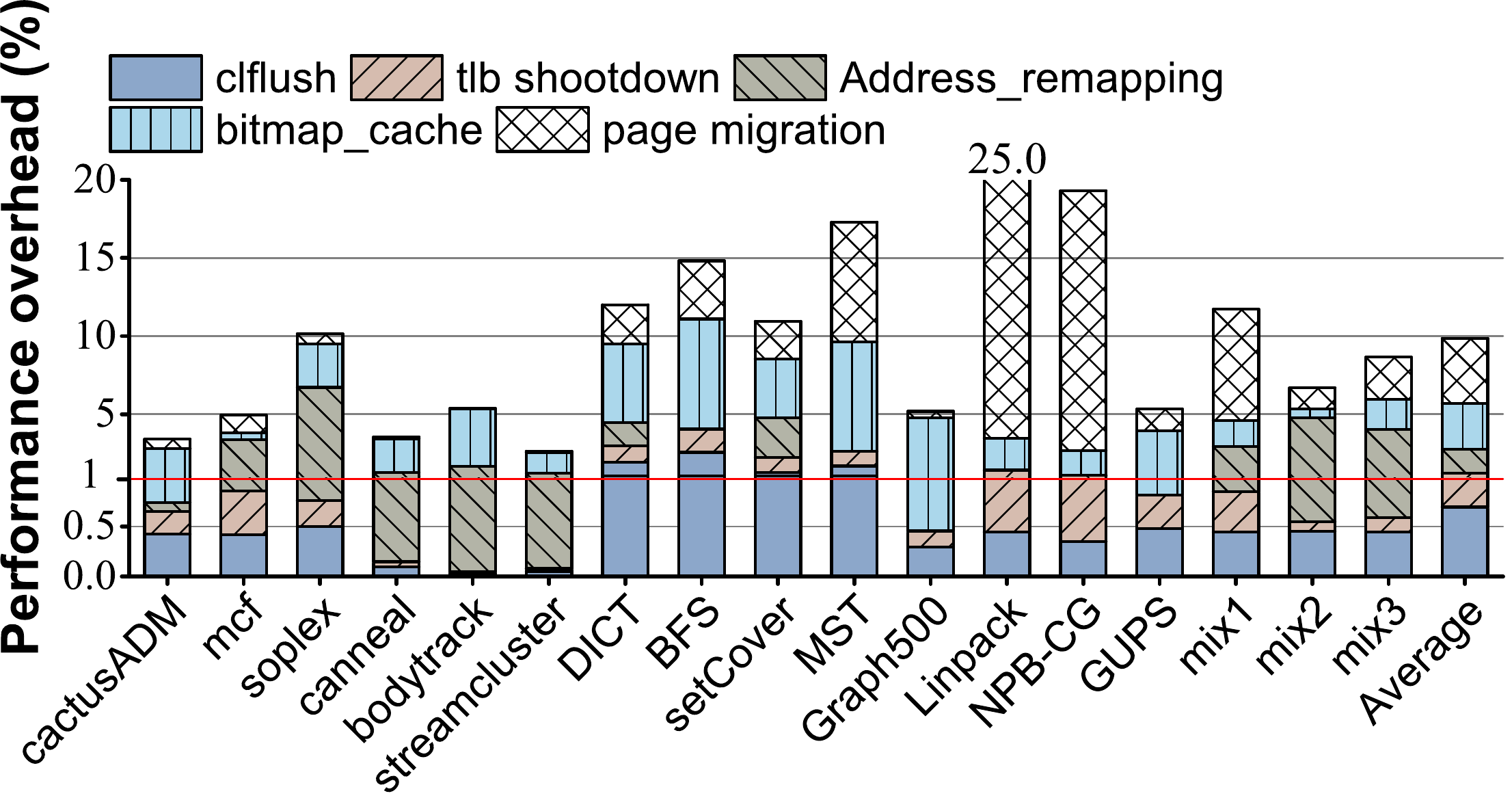}
    \vspace{-3.0ex}
	\caption{Breakdown of runtime overhead in Rainbow}
	\label{fig:runtimeoverhead}
	\vspace{-3ex}
\end{figure}

Figure~\ref{fig:runtimeoverhead} shows the breakdown of performance overhead due
to the address remapping mechanism, bitmap cache, page migration, TLB shootdown, and clflush. We model all these operations in our simulator by adding reasonable latencies accordingly. We find that these applications show significantly different performance overhead at runtime. For soplex, mix2 and mix3, DRAM page addressing accounts for the majority of runtime overhead, implying that these workloads show relatively high miss rate of 4 KB page TLBs. DICT, BFS, setCover, MST and Graph500 all spend a relatively large portion of time in accessing the bitmap cache, suggesting that many memory accesses are distributed on NVM due to poor data  locality of these workloads. MST, Linpack, and NPB-CG show very large memory footprints, and thus a large fraction of execution time are spent on page migrations. Overall, the runtime performance overhead of Rainbow is 9.8\% on average. However, it can be offset by the significant benefit of using superpages and lightweight page migrations.

\section{Related Work} \label{relatedwork}
We summarize the related work in the following categories.

\textbf{Superpages and TLBs}.  There have been many studies on mitigating the performance overhead of virtual-to-physical address translations~\cite{basu2013efficient,pham2012colt,Bhattacharjee2010Inter-core,bhattacharjee2011shared, srikantaiah2010synergistic}. Due to energy and latency constraints on TLB designs, a majority of studies have focused on superpages for improving TLB coverage. Talluri \textit{et al.} has discussed the challenges and tradeoffs to support superpages in hardware~\cite{Talluri1992tradeoffs}. Libhugetlbfs~\cite{libhugetlbfs} and Ingens~\cite{Kwon:2016} are OS-level supports for hugepage management. TLB coalescing~\cite{pham2012colt, pham2014increasing} and MMU cache coalescing~\cite{Bhattacharjee:2013:LMM:2540708.2540741} have been proposed to increase the coverage of TLB and MMU cache. GLUE~\cite{Pham:2015} groups contiguous, aligned small page translations under a single speculative huge page translation in the TLB. Redundant memory mappings (RMM)~\cite{Karakostas2015RMM} extends TLB coverage by mapping ranges of virtually and physically contiguous pages in a range TLB.

Many studies have focused on improving the availability of superpages. Navarro \textit{et al.} propose reservation-based allocation and deferring promotion~\cite{Navarro2002practical} to support superpages in the OS layer. Gorman \textit{et al.}~\cite{gorman2008supporting} propose a physical page allocator to mitigate memory fragmentation and promote page contiguity. Zhang \textit{et al.} proposed Enigma to map superpages to discontinuous physical pages using a intermediate address (IA) space~\cite{zhang2010enigma}. GTSM~\cite{du2015supporting} leverages contiguity of physical memory extents to construct superpages even when pages have been retired due to bit errors. MIX TLB~\cite{Cox:2017} exploits superpage allocation patterns to concurrently support multiple page sizes. Those studies are orthogonal to our work as the design space is different. \textit{Rainbow} mainly aims to address a thorny problem of enabling lightweight page migration in a superpage-supported hybrid memory system.

\textbf{Page Migration in Hybrid Memory Systems}. There have been a number of studies on page migration for hybrid memory systems~\cite{dhiman2009pdram, ramos2011page,Liu:2017,DI-MMAP}. Both PDRAM~\cite{dhiman2009pdram} and CLOCK-DWF~\cite{lee2014clock} migrate frequently-written NVM pages to DRAM, while remaining read-intensive pages in the NVM. AIMR~\cite{AIMR2015} exploits data write ``recency" and ``frequency" to identify write-intensive pages, and only migrates these NVM pages to DRAM.
RaPP~\cite{ramos2011page} ranks pages according to the access frequency and recency, and migrate the top-ranked NVM pages to DRAM. HSCC~\cite{Liu:2017} extends the TLB and page tables to count NVM page accesses, and explores an utility-based model to migrate hot NVM pages to DRAM. Bock \textit{et al.} proposed CMMP~\cite{bock2016concurrent} to concurrently migrate multiple pages. Those studies have assumed that the hybrid main memories are uniformly managed at the granularity of 4 KB pages, and thus naturally supports lightweight page migration~\cite{Qureshi:2009}. The context of \textit{Rainbow} is different from those studies. Rainbow mainly focuses on supporting lightweight page migration in hybrid memory systems while still preserving the benefit of superpages.

Probably the most relevant work to this paper, Thermostat~\cite{Agarwal:2017} supports page migration at the granularity of 2 MB or 4 KB pages for a two-tiered hybrid memory system. Rainbow is different from Thermostat in two folds. First, to migrate small pages, Thermostat needs to splinter the corresponding superpages and then migrates the small pages. In contrast, Rainbow supports lightweight page migration without splintering superpages, and thus preserves the benefit of superpages on TLB performance. Second, Thermostat exploits an OS-level extension for intercepting TLB misses to estimate access counts at the 4 KB page granularity. The software overhead is usually rather high, and thus Thermostat make a tradeoff between the precision of hot page monitoring and the performance penalty. In contrast, the two-stage page access counting mechanism in Rainbow is more precise and efficient than thermostat, and thus leads to lower page migration cost.

\vspace{-0.5ex}
\section{Conclusion}\label{sec:conc}
Superpages are able to significantly improve TLB coverage and reduce address translation overhead. Hybrid memory systems composed of DRAM and NVM usually can provide very large memory capacity, and thus are more eager for the support of superpages. However, superpages often preclude lightweight page migration, which is a key technique for improving performance and energy efficiency in hybrid memory systems.
In this paper, we propose a novel hybrid memory management mechanism called \textit{Rainbow} to support both superpages and lightweight page migration. \textit{Rainbow} manages the NVM at the superpage granularity, and uses the DRAM to cache frequently-accessed (hot) small pages in the superpages.
Correspondingly, Rainbow utilizes split TLBs to support different page sizes. We propose a NVM-to-DRAM address remapping mechanism to identify the migrated small pages, without splintering the superpages. Experimental results show that Rainbow can significantly reduce the address translation overhead for applications with large memory footprints, and improve application performance by up to 2.9X (43.0\% on average) compared to a state-of-the-art memory migration policy without superpage support.

\vspace{-0.5ex}
\bibliographystyle{unsrt}
\bibliography{ref}

\end{document}